\documentclass[fleqn,9pt]{manuscript}
\usepackage{soul}
\usepackage{multirow}
\usepackage{url}
\usepackage{pdfpages}

\title{Social bots sour activist sentiment without eroding engagement}

\author[a]{Linda Li}
\author[b,c,d]{Orsolya V\'as\'arhelyi} 
\author[e,a,b]{Bal\'azs Vedres}

\affil[a]{Oxford Internet Institute, University of Oxford, Oxford, UK}
\affil[b]{Center for Collective Learning, Corvinus Institute for Advanced Studies, Corvinus University, Budapest, Hungary}
\affil[c]{Institute of Data Analytics and Information Systems, Corvinus University, Budapest, Hungary}
\affil[d]{CEU Democracy Institute, Budapest, Hungary}
\affil[e]{Department of Network and Data Science, Central European University, Vienna, Austria}

\corrauthor[e,a,b]{Bal\'azs Vedres}{vedresb@ceu.edu}

\keywords{social bots; human-bot interaction; information cascades;  political communication; protests}

\begin{abstract}
Social media platforms have witnessed a substantial increase in social bot activity, significantly affecting online discourse. Our study explores the dynamic nature of bot engagement related to Extinction Rebellion climate change protests from 18 November 2019 to 10 December 2019. We find that bots exert a greater influence on human behavior than vice versa during heated online periods. To assess the causal impact of human-bot communication, we compared communication histories between human users who directly interacted with bots and matched human users who did not. Our findings demonstrate a consistent negative impact of bot interactions on subsequent human sentiment, with exposed users displaying significantly more negative sentiment than their counterparts. Furthermore, the nature of bot interaction influences human tweeting activity and the sentiment towards protests. Political astroturfing bots increase activity, whereas other bots decrease it. Sentiment changes towards protests depend on the user's original support level, indicating targeted manipulation. However, bot interactions do not change activists' engagement towards protests. Despite the seemingly minor impact of individual bot encounters, the cumulative effect is profound due to the large volume of bot communication.
Our findings underscore the importance of unrestricted access to social media data for studying the prevalence and influence of social bots, as with new technological advancements distinguishing between bots and humans becomes nearly impossible. 
\end{abstract}

\begin{document}

\flushbottom
\maketitle
\thispagestyle{empty}

\section{Introduction}
Social media has become the primary channel to learn about and spread news, engage in political discussions, and participate in collective actions over the past decade \cite{Farrell2012,Keller2020,Caren2020}. Automated agents, or so-called "social bots," are increasingly active in our social and political publics online \cite{Hepp2020b}, creating content and interacting with humans \cite{Seering2018}. The resulting hybrid ecosystem where algorithmic agents and humans co-exist can fundamentally alter the nature of democracy, political accountability, transparency, and civic participation \cite{Rahwan2019a}. Automated accounts can propagate a large volume of messages at minimal expense, and engage with users promptly and excessively, far exceeding human capacities \cite{Assenmacher2020a}. As a consequence, the algorithmic share of social media content is now on par with human participation: Automated users were estimated to be responsible for generating 10\% to 40\% of tweets in recent political events,  such as the 2016 US presidential election, the Brexit referendum, the yellow vests movement, the Catalan referendum, or the 2019 United Nations Climate Change Conference \cite{Bastos2019, Keller2019, Gonzalez-Bailon2021}.

Recent studies have found that social bots -- especially those designed to mimic human behavior --, disrupt online political discussions \cite{Woolley2016}, and there is overwhelming evidence that they significantly influence political debates and activism\cite{Woolley2016, Ferrara2016, Ferrara2018}. Such bots are frequently designed to pass as human accounts, and occasionally explicitly mimic known political figures and government accounts to gain the attention and trust of human users \cite{Forelle2015}. Bots are often highly active during the flare-up of discussions around new political events \cite{Shao2018c}, and disseminate targeted messages ranging from fabricated news to contentious, divisive, and negative content\cite{Stella2018, Grinberg2019, Vosoughi2018a}, blending legitimate messages and misinformation. Furthermore, bots are also often deployed in orchestrated efforts to generate the facade of a seemingly vibrant discussion conforming with hidden agendas \cite{Stella2019a, 10.5555/3133434} by retweeting each other (known as "astroturfing"), targeting susceptible users\cite{Bastos2019}.

Research on bot-human interaction found that bots can often hijack the collective attention of human users \cite{Gonzalez-Bailon2021, Stella2018}, can increase the visibility of extreme views \cite{Bastos2019, Stella2018, Schafer2017}, influence communication sentiment \cite{Hagen2020}, and can even alter human communication networks \cite{Oliveira2016, Kobis2021}.  Although social bot presence has been studied before at the macro scale, less is known about the micro-level impact of human-bot encounters on subsequent human activism. Small-scale simulations and experiments indicate that bots can alter expressed human values \cite{Kobis2021} and behaviors \cite{Mosleh2021}, particularly driving users towards more extreme viewpoints \cite{Bail2018b} and potentially silencing them \cite{Ross2019}. However, there is little empirical evidence regarding the capacity of bots to modify human behavior in real-life political communication\cite{Grinberg2019, Stella2019a, Oliveira2016,Schuchard2019}. 

Today, online activism constitutes an essential part of democracy, but most of the existing research focuses on institutionalized political processes such as elections \cite{Hagen2020, Stukal2019, Murthy2016, Schafer2017}, while few studies examined the role of bots within activism related to online protests \cite{Gonzalez-Bailon2021, Oliveira2016, Salge2018}. Furthermore, these studies have often collapsed acts of communication, analyzing them in aggregate. The dynamics of social media use in online activism differs markedly from communication around other political events \cite{Gonzalez-Bailon2013, Earl2022}, and is also a critical component of online political discourse \cite{Freelon2018, Jennings2019}. Since social bots are designed to be more responsive than humans \cite{gilani2019large}, it is common to observe an increase in bot activity during the peak of heated online debates, followed by a decrease in their presence\cite{Bastos2019}. Although the bursty nature of social media communication of protests is known \cite{shirakawa-etal-2017-never, comito2019bursty}, the impact of social bots on human activity during and after bursty periods has not been investigated.

How does interacting with social bots impact human behavior in online activism? We address this question with data on Twitter discourse on climate-change-related social movements. We analyze the dynamics of bots and humans engaging with each other, and we also compare the difference in impact of bot encounters to activists who did not engage with bots. Our analysis focuses on protest-related discourse during a series of protest events that erupted from November to December 2019. We decided to concentrate on online activism related to climate change as our case study, as algorithmic threats to engagement in climate change activism can have profound consequences on societal agreement on the public good in a critical issue \cite{Carter_2018}. We analyze communication around the Extinction Rebellion (XR), as the highest profile activist group online. 

The topic of climate change has been shown to attract highly engaged, active, and committed participants \cite{Bennett2012a}, while there is also substantial bot activity \cite{marlow2021, Chen2021}. This enables us to analyze the impact of human-bot interactions on humans during information cascades \cite{Stella2019a}, and measure the effect of bot encounters on tweeting activity\cite{Ross2019} and sentiment \cite{Bail2018b}. Our findings contribute to the growing body of research on machine behaviour\cite{Rahwan2019a}, particularly in terms of understanding how rapidly developing hybrid human-machine systems could potentially modify human opinion over an extended period.

Early work related to human-bot interactions on social media found that user sociability and network size predict who will be interacted with by a bot \cite{wagner2012social, Shao2018c}. Our results show that bots have become so widespread that it is unlikely that an active user of X will {\it not} interact with a bot online. Our research extends previous work by focusing on quantifying the impact of direct human-bot communication. We found that bot type matters for the impact on users' tweeting activity, and the initial level of support a user has toward the climate change movement determines how a bot encounter impacts their sentiment on climate change. Our results have important policy implications for increasing platform transparency in how they handle automated profiles. 

\section*{Results}

\subsection*{Proportions of bot and human communication} 

%Bots retweet humans significantly less ($\chi^2=27.16$,$p<0.001$) than humans retweet bots, which makes it harder for activists to tackle the challenges that automated targeted harmful campaigns cause \cite{Greve2022, Keller2019, Schafer2017}. 

We found $48\%$ of all accounts to fall into the bot category within our sample from Twitter ($44,121$ of the total $93,499$ users). We identified automation through a combined approach \cite{alothali2018detecting,martini2021bot} that integrates the results of commonly used bot identification methods, Botometer\cite{DBLP:journals/corr/abs-2201-01608} and our self-trained machine learning-based algorithms. Botometer is a widely adopted open source tool to identify bots on Twitter \cite{10.1371/journal.pone.0241045}. Our self-trained models used various data sets tailored to identify users with automated behavior that attempts to mimic humans on social networks, especially in the context of political behavior \cite{DBLP:journals/corr/abs-2201-01608}. Bots reported in the main text are the combined results of these two models, with a fixed threshold for the Botometer ($CAP>=0.65$). Since the concept of "bot" encompasses varying degrees of automation, using one fixed bot classification threshold is always a simplification. Therefore, we repeated all analyzes reported in the main text at various bot thresholds, and we report these results in the SI.(See Materials \& Methods, Bot Identification and SI Bot Identification for more details). 

81\% of tweets were replies or retweets in our database, and 51\% of these retweets originated from bots. In general, bot activity is mainly the posting of original messages or the retweeting of each other (35\% of all retweets and 71\% of bot retweets were retweeting other bots), and only a small portion of bot retweets were retweets originating from humans (29\%). At the same time, humans spread roughly the same number of messages produced by bots (54\%) and humans (46\% ). If the bot threshold is increased to $CAP>=0.75$, still 48\% of the retweets originate from bots, and 45\% of the human retweets originate from bots (see SI Information cascades and SI Tables S5 to S7 for information flows). 

\begin{figure}%[tbhp]
\centering
\includegraphics[width=.8\linewidth]{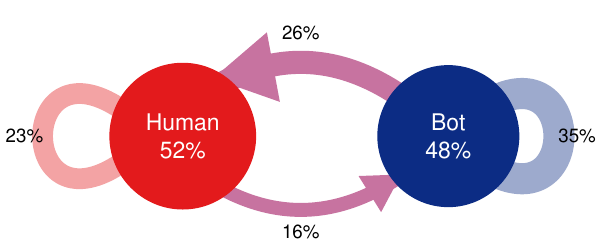}
\caption{Information flow of humans and bots. The directions and amount of tweets retweeted between bots and humans.}
\label{fig:infromation_flow}
\end{figure}

\subsection*{Temporal co-dependence of bot and human tweeting} 
We found that the quantity, intensity, direction and the sentiment of the information flow between humans and bots are highly topic-dependent. We identified seven topics related to XR protests associated with news events, political campaigns, or outbursts of sentiments (for example, climate change denial) by applying bi-term topic models \cite{Yan2013}. (See Materials \& Methods, Topic Modeling for more details and SI Table S4 for a brief summary of the topics, including their content, top hashtags, and sample tweets). Out of these seven topics, four were highly bursty: Topic-related activity increased sharply periodically and then decreased suddenly\cite{goh2006burstiness}, resulting in information cascades. Information cascades occur when users follow the behavior of other individuals on social networks (e.g. retweeting the same message) \cite{Easley2010}. (See Table S8 in Supplementary Information on Burstiness Scores).

Figure \ref{topic-cascade} shows the temporal distribution of human and bot tweets at a 5-minute interval within an illustrative bursty period of climate change discussions on Twitter. Although human users (red line) generated a higher volume of tweets during the cascade's peak, the tweeting frequency trend is jointly influenced by both bots and humans.  Table \ref{tab:granger-casuality} shows the results of Granger Causality tests, which examine whether one time series could be used to predict another after introducing a time lag \cite{Hsiao1979}. 
Our results indicate that in all identified cascades, the number of bot tweets during bursty periods could be used to predict human activity. Also, in $3$ cases, the number of tweets by humans could be used to predict bots' activities, suggesting that the effect was mutually driven, indicting that bots and humans drove the cascade together. The effect was stronger on bots' side in two of those three mutually-driven cascades, and in $3$ out of $4$ cascades humans retweeted bots much more than bots retweeted humans. Furthermore, in $3$ out of $4$ cascades, the sentiment of bot tweets could be used to predict human sentiment after introducing a 30- or 35-minute time lag. The effect was mutual only for 'Anti XR protests', but the effect was stronger on bots' side. This suggests that bots not only capture human attention, but also have an immediate impact on the sentiment of humans exposed to them. The effect remains consistent across bot CAP thresholds ranging from $0.50$ to $0.75$ for $3$ of the identified $4$ cascades.  (See more details on Burstiness in Materials \& Methods, Identifying cascades, SI Cascades, SI Table S8-S10 for Granger Causality Results across different CAP thresholds and for Topic Burstiness Scores).

\begin{figure}%[tbhp]
\centering
\includegraphics[width=1\linewidth]{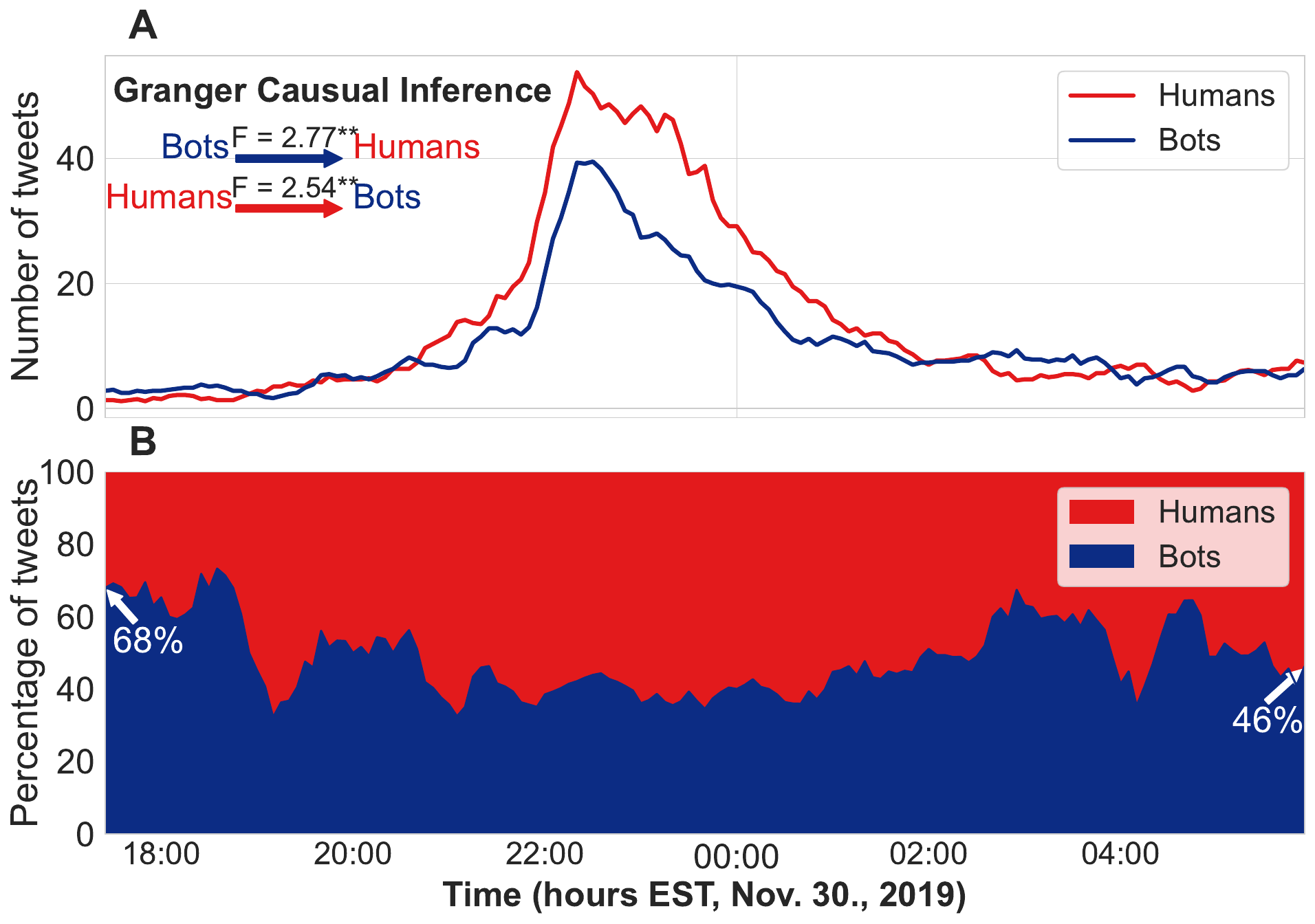}
\caption{Panel A: Number of tweets posted by bots (blue line) and humans (red line) in a cascade mutually-driven by bots and humans within the "Anti-XR protests" topic.  Panel B: Ratio of bot and human generated tweets throughout the time period of the cascade. The number of tweets is the rolling average aggregated on 5-minute  intervals.} %See more details on Burstiness in Materials \& Methods, Topic Burstiness and SI Table S8-S10 for Topic Burstiness Scores.) Area chart (b) shows the ratio of human and bot-generated tweets during bursty time periods.

%We calculate the percentage of tweets that belonged to each topic for every hour, and then calculate the z-score of all those percentages to standardize them. A period is bursty for a topic, when the z score of a topic in proportion among all topics during the given hour is greater than $2$ for two consecutive hours, and the total number of tweets in the topic in an hour is greater than $50$. }
\label{topic-cascade}
\end{figure}

\begin{table}[]
\begin{tabular}{rrlrlrlrl}
\toprule
\multicolumn{1}{c}{\multirow{2}{*}{Topic}}                                                                     & \multicolumn{4}{c}{Bots$\rightarrow$Humans}                                                                                                             & \multicolumn{4}{c}{Humans$\rightarrow$Bots}                                                                                                             \\
\multicolumn{1}{l}{}                                                         & \multicolumn{2}{c}{\begin{tabular}[c]{@{}c@{}}Amount\end{tabular}} & \multicolumn{2}{c}{\begin{tabular}[c]{@{}c@{}} Sentiment\end{tabular}} & \multicolumn{2}{c}{\begin{tabular}[c]{@{}c@{}}Amount\end{tabular}} & \multicolumn{2}{c}{\begin{tabular}[c]{@{}c@{}}Sentiment\end{tabular}} \\
\midrule
\begin{tabular}[c]{@{}r@{}}"Football game\\ protests"\\\end{tabular}             & 2.27                                  & *                                  & 0.38                                  &                                    & 3.97                                  & ***                                & 0.92                                  &                                    \\
\begin{tabular}[c]{@{}r@{}}"Disruptive\\engagement"\end{tabular} & 15.60                                 & ***                                & 2.17                                  & *                                  & 2.68                                  & **                                 & 1.91                                  &                                    \\
\begin{tabular}[c]{@{}r@{}}"Anti-XR \\ protests"\end{tabular}                  & 2.77                                  & **                                 & 3.96                                  & ***                                & 2.54                                  & **                                 & 3.08                                  & **                                 \\
\begin{tabular}[c]{@{}r@{}}"Politicized\\activism"\end{tabular}               & 6.06                                  & ***                                & 11.40                                 & ***                                & 1.50                                  &                                    & 2.24                                  &                           \\                     
\bottomrule
\end{tabular}
\caption{Granger causality test results between bot and human communication amount and sentiment. *: $p<0.05$, **: $p<0.01$, ***:$p<0.001$. Granger causality tests whether bot activities can predict human activities (bot-human) or human activities can predict bot activities (human-bot) in all four identified cascades. These tests were performed on the first-order differences for both the number of tweets (aggregated in 5-minute intervals) and the average sentiment (calculated over 5-minute intervals). The time lags used in the tests were set at $30$ or $35$ minutes.} %For additional details and results, please refer to SI Granger Causality and SI Tables S8-S10}
\label{tab:granger-casuality}
\end{table}

\subsection*{Predicting the impact of direct bot interactions} 
%Existing research has shown that engagement in online debates (activity) and the emotional intensity of involvement (sentiment) can significantly influence the dynamics of political communication and contentious political processes \cite{Dang-Xuan2013a, Ceron2014,Zerback2017}. 
To test whether bot encounters have any influence on human users - beyond their above shown immediate collective  effect in bursty periods - we developed two models focusing on 1) Amount and 2) Sentiment. Specifically, we analyzed how human communication evolves over a span of $30$ days after the first interactions with a bot in the discussion related to climate change on Twitter. The first model captures the inclination to "speak out" quantified by the average number of tweets posted related to the XR protest. The second model predicts the change in sentiment about the climate change protest (measured on a scale ranging from $-1$ to $1$, with $-1$ representing the most negative sentiment and 1 the most extreme sentiment). 

\begin{figure}
\centering
\includegraphics[width=1\linewidth]{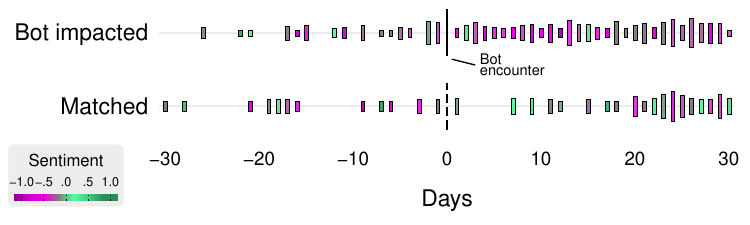}
\caption{Two examples of Twitter timelines. Top timeline shows daily tweet counts and daily mean sentiment for a human account that directly interacted with a bot on day 0; bottom timeline shows a matched account without a direct bot interaction. Bar heights are proportional to tweet counts, color indicates sentiment from $-1$ to $1$. }
\label{fig:timeline}
\end{figure}

To quantify the impact of bot interaction on human communication change, we used a three-step process. This includes: 1) we sampled $N=303$ users who directly interacted with bot accounts (bot-exposed) - replied or commented to a tweet/comment originated from a bot; 2) we collected a matched sample of $N=184$ users, who were active in the XR related protest discussion on Twitter but had no direct interaction with bots. Matching users were selected on the basis of the similarity score calculated pairwise to our original sample considering publicly available metrics on human users' profiles. (See Materials and Methods, Matching Sample for more details.) 3) Figure \ref{fig:timeline} shows two example timelines, one human user who directly interacted with a bot (the interaction is shown at time 0), and a matched user that did not encounter a bot.

3) Finally, we applied Difference-in-Difference (DiD) \cite{athey2006identification} regression models to quantify the casual effect of meeting a bot on outcomes by comparing a set of humans who directly interacted with bots (human replying or commenting bot tweets) with those who did not. Our observation units are the daily activity of human users relative to the time of interaction with bots - $30$ days {\it before} and $30$ days {\it after}. This setting allows us to quantify the prolonged impact of bot interaction on the frequency of tweeting and sentiment of tweets compared to users who did not meet a bot.

\begin{figure}
\centering
\includegraphics[width=1\linewidth]{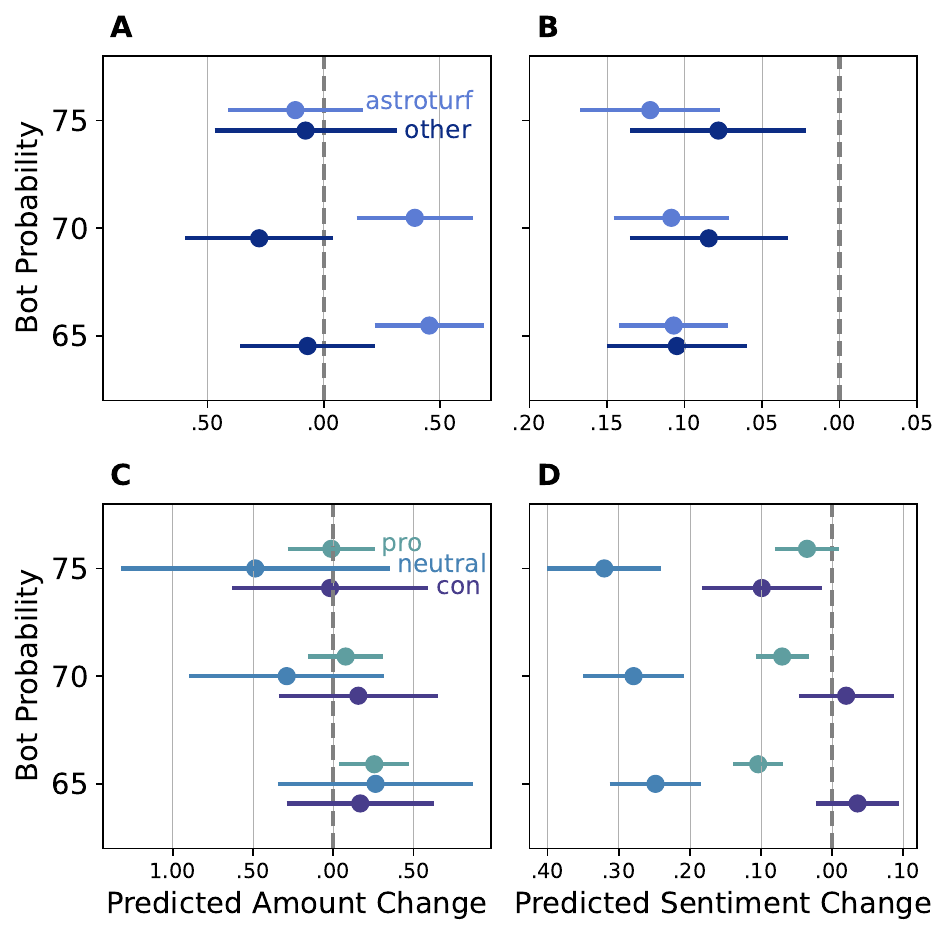}
\caption{Predicted change resulting from a bot interaction. Panel A, C: Predicted change in Amount (number of tweets). Panel B,D: Predicted change in Sentiment. Panel A-B separate predictions (by color) are shown for astroturf bots (light blue), and other bots (dark blue), while Panel C-D separate predictions by users support level towards XR - supporters (green), neutral users (blue), anti-XR (purple)}
\label{fig:pred}
\end{figure}

Botometer provides prediction values for belonging to seven subbot categories ranging from $0$ to $5$. We classify a bot into a subcategory if its bot-specific probability is greater than $2.5$. Our biggest group is 'miscellaneous other bots' which are bots that are similar to various type of manually annotated bots (23\%), followed by manually labeled political bots, so-called 'astroturfs' (14\%),'fake followers' bots purchased to increase follower counts (7\%), 'self declared' bots  from botwiki.org (5\%), 'spammers' (2\%) and 'financial bots' (0.4\%). Although 'miscellaneous other bots' are quite well represented within the XR discourse online, only 17\% of human users had direct interaction with them. As implied by the topic of our analysis, the individuals in our sample primarily engaged with 'astrotufs' (38\%). These automated accounts are specifically designed to participate in political discussions \cite{Keller2020}, leading us to analyze users interacting with  astroturfs separately from those who interacted with non-astroturfs.

Figure \ref{fig:pred}. visualizes the predicted daily change in the average number of daily tweet counts (Panel A), and the sentiment of the tweets (Panel B) grouped by different bot probabilities (65,70,75) and subbot category (astroturf or other type of bots). We found that encounters with astroturf bots result in an increase in the number of tweets, while encounters with other kind of bots result in a decrease in the number of tweets. This suggests that the most politically relevant bot category - astroturf bots - drive the conversation by provoking engagement from human users, while other kinds of bots have a rather negative, silencing effect. Regardless of the bot type, direct interaction with a bot decreases the average sentiment of human users. (See Table S12-S17 for DID models)

Astroturfs tend to be mobilized in a targeted way against users with a specific opinion\cite{Shao2018c, varol2017online}. Therefore, we classified bot-exposed and matched human users by their support of XR Protests: Supporters (Pro - 52\%), Neutral (27\%), and Anti-XR (Con - 21\%) using ChatGPT. (See Materials and Methods on Support Categorization). Figure \ref{fig:pred}, panel C-D indicates that human users who support or have a neutral opinion about XR are significantly affected by interacting with bots, while anti-XR users are not affected. Bot interaction has the strongest negative impact on the sentiment of bot-exposed users with neutral opinion, indicating that bots might target those users whose opinion can be changed \cite{Shao2018c, Varol2017}. The change in the number of tweets was significantly increased by bot interaction for XR supporters within the least selective bot probability category (65), although the trends are similar in the more selective categories (.70,.75). In these models, we control for the astroturf score of the interacted bot, which has a strong positive significant relationship with the change in amount. Our results indicate that the type of bot matters more for the activity change than the original support towards protests. However, the level of sentiment change depends on the original support level of users exposed to bots.  (See SI Table S18-S23 for DID models) 

We also investigated if users affected by bots alter their levels of support as a result of bot interaction. There were no significant differences between the distribution of bot-exposed users' support level before, and after interacting with a bot (Mann-Whitney U=45536.00, p=0.86).There was no significant difference either between users exposed to bots and those matched in terms of the change in support level.(See Model Tables 23-24 in SI on Opinion Change). Out of the users who interacted with bots, only 9\% experienced a shift towards neutrality, while 10\% of the opinions of the users in the matched group became more neutral towards XR protests. Additionally, 7\% of the users exposed to bots and 6\% of the matched users changed their opinion completely, shifting from negative to positive or the other way around.

\section*{Limitations}
There are two potential limitations to our current research design. The first is the relatively short time range and sample size of our dataset on XR protests. Our dataset covers several waves of XR protests within a month, but XR is a global phenomenon that has been going on for several years. However, comparing our sample size to previous studies on online activism and political communications on Twitter\cite{Freelon2018,Gonzalez-Bailon2013}, we believe that it is a valid and representative case to illustrate the studied aspects of bot activities on online activism. The Twitter academic product track API (available at the time of data collection) provided the full archive of tweets based on specific search queries; therefore, our dataset is a comprehensive sample of the online record of bot and human activities during the protest period. 

The second limitation concerns our binary bot detection method. It has three main drawbacks, each of which we address through additional efforts and measures. 
First, similar to other bot identification approaches \cite{DBLP:journals/corr/abs-2201-01608, DBLP:journals/corr/abs-1901-00912, Gonzalez-Bailon2021, Feng2022}, and due to the nature of unsupervised learning, our method cannot be a 100\% sure whether a user classified as a bot is genuinely a bot. 
Second, the concept of "bot" encompasses varying degrees of automation among Twitter users, and using a fixed bot classification threshold overlooks these nuances. We are aware that being a bot, similar to the concept of gender \cite{vedres2019gendered}, is not a binary classification problem. Many human users apply automation to increase their efficiency \cite{Chu2012}, which does not turn them into bots, but they are no longer non-automated human users. 
Third, a fixed threshold-based approach can still lead to false positives and false negatives, potentially undermining the validity of our causal inferences regarding bot activities. To address those concerns, we ran our models with varied CAP scores. We also performed a series of robustness checks involving various thresholds for all of our analysis and reported them in the main text. (See Discussions, SI Information flow, Cascades, and Difference-in-Difference regression results).

\section*{Discussion}
Bot presence is considerable and possibly increasing in the public sphere. Even with stricter thresholds, bot activity is higher in our sample than in previous related work published on the 2017 Catalan referendum by Stella et al. \cite{Stella2018}, where only 19\% of all interactions were from bots to humans. It was shown that tweets from conservative bots are more retweeted by humans, which can indicate that this difference is not only due to the continuously increasing presence of bots \cite{luceri2019red}, but could be due to the highly politicized and international nature of our context. 

We have adopted a dynamic approach to political communication, and we have found that bots should not be thought of a constant presence, but rather bots react in a dynamic fashion, and are driving the amount and sentiment of heated and bursty discussions. By our analysis of Granger causality, bots impact human behavior more, than the other way around. 

To quantify the causal impact of human-bot communication, we have compared the communication histories of human users who have directly interacted with a bot with those matched human users who have not. This allowed us to see a consistently negative impact of any kind of bot interaction on the sentiment of subsequent human communications: Humans who have interacted with a bot displayed considerably more negative sentiment than matched users. 

The change in tweeting activity after a bot interaction depends on the nature of the bot: On one hand, decidedly political astroturfing bots (aiming to influence public opinion behind an impression of grassroots opinion) result in an increased activity. On the other hand, interacting with other kinds of bots (spammers, fake followers) results in a decrease in activity. However, change in sentiment towards the protests depends on the original support level of the user, supporting, neutral, or against XR. Our results indicate that bots might target users whose opinions are easier to change, since the sentiment of bot-exposed users with neutral opinion decreased the most. However, bots do not make human users switch their support level. In sum, bot interaction is not without impact, even if one encounter itself has only a small effect (it takes about two bot encoutners to induce one additional tweet). Nevertheless, since there is an exceeding amount of bot communication, these small impacts add up to influence the public sphere in a profound way.

Although our analysis covers a period of time prior to the launch of ChatGPT (2022.11.30), it is becoming increasingly difficult to identify bots due to the rapid advancement mimicking human behavior\cite{mei2024turing}. As large language models are becoming widespread and easily accessible through APIs, new social bots can act extremely human, making it currently almost impossible to distinguish between bots and humans, even for experts \cite{jo2023promise, yang2023anatomy}.

Therefore, it is crucial to have unrestricted access to social media data to assess the influence and prevalence of these new types of social bots, although recent trends show that social media platforms are less willing to share free data for research purposes \cite{verge_twitter_musk, verge_reddit_api_changes}. Since financial evaluations are highly correlated with the size of the (human) user base \cite{techcrunch_irl_shutdown}, platforms have no interest in quantifying the ratio of non-human accounts and their impact on misinformation \cite{Shao2018c}. Most users still underestimate the effect of bots on themselves, but as they are exposed to increased bot presence, they tend to prefer stricter bot-regulation policies  \cite{yan2023exposure}. That is why we welcome the news that the European Union requires larger platforms to provide researchers with access to data to study systemic risks arising from the use of their services, such as disinformation \cite{eu_digital_services_act}. Such legislative actions can help the scientific community continue its work to understand the consequences of this abrupt change in technology that will alter the nature of human-bot interactions \cite{jo2023promise, 10.1145/3593013.3593981, li2023you, del2023large}.

\section*{Data and Methods}
\subsection*{Data}

Our data is made up of Twitter activity around several waves of Extinction Rebellion climate change protests from 18 November 2019 to 10 December 2019. The dataset was collected from November to December 2022 via the Academic Research product track API provided by Twitter\cite{Twitter2022}, which enabled users to collect a full archive sample of historical tweets filtered based on keywords and conditions. We collected all tweets posted during this period of time that contained the keyword 'Extinction Rebellion', 'climate change protest', 'XRebellion', 'XR' and multiple variants of keywords with slightly different spelling. (The complete list of keywords used can be found in SI Table S1) In total, the final data set contained 201,010 tweets and 122,130 users. 

\subsection*{Bot Identification}
To identify social bots on Twitter, we used a combination of two sets of bot identification methods. The first is a popular Twitter bot identification tool known as the 'botometer' (formerly BotorNot), which was primarly used as a benchmark to compare other methods. The second is a set of our self-trained bot identification model trained with open source data of bots and humans to train supervised machine learning models.

Botometer is a publicly available tool that relies on machine learning. It is designed to calculate a score where low scores indicate likely human accounts, while high scores suggest likely bot accounts \cite{DBLP:journals/corr/abs-2006-06867}. The algorithm considers more than 1,000 features related to user profiles, friends, network structure, and activity patterns, among others. Another part of our bot identification pipeline comes from self-trained models. Training sets were derived from existing open-source data from Twitter accounts identified as 'bots' and 'humans'.  

We trained bot identification models with 70\% training and 30\% of testing set with five types of algorithms:  random forest(RF), support vector machine(SVM), logistic regression(LOG), XGboost classification (XGB) and deep learning (DL). We developed two versions of our mdoels with ten and twenty features that were proved to be most effective for bot identification by previous studies \cite{Yang2020, Gonzalez-Bailon2021}. The evaluation of the models demonstrated that the RF, DL, and XGB models with 20 traits surpassed other models in terms of sensitivity (true positive rate), balanced precision, precision, and F1 score. Additionally, these models exhibited strong performance in an independent test carried out on a dataset consisting of influential bots and human mimics that were active during the 2018 midterm election of the United States.

In our final bot identification approach, we combined the results derived from both sets of bot identification methods. Due to the potential false positive issues inherent in both methods, they yielded somewhat divergent results \cite{10.1371/journal.pone.0241045}. To reconcile this disparity, we classified users based on the overlapping results of botometer {\it and} our proprietary algorithms (DL, RF, or XGB). If both the botometer and at least one of our algorithms identified a user as a bot, the user was classified as such; conversely, the same principle was applied to the classification of humans. To account for the potential error in bot identification, we performed all our analyzes with a varying baseline CAP score ranging between $0.60-0.75$ (See more details in Supplementary Information, Bot Identification).

\subsection*{Topic Modeling}
We first identified the themes emerging in the protest-related discourse with the increasingly popular bi-term topic models \cite{Yan2013, Shi2019} that learn topics by modeling word-word co-occurrence patterns \cite{BTMR}. After removing all retweets, to preprocess the data we used the nltk python package \cite{bird2009natural} to remove stopwords, usernames, emojis and links from the tweets, and lemmtize and stemm every word.

We then trained bi-term topic models on our preprocessed data with the bitermplus package\cite{Terpilovskii2022}. We set up a biterm topic model of all the tweets related to XR, which classified all tweets into 8 different topics. Based on the u\_mass coherence scores \cite{10.1145/2684822.2685324}, we determined that 8 topics fits our dataset the best. After evaluating the meaning of topics we dropped one topic whose keywords and content were too diverse to extract a meaningful media agenda from it.
(See more details in Supplementary Information, Topic Modeling).

\subsection*{Granger Causality}
We tested whether bots attracted people’s attention to given topics, through 'granger causality' tests. Granger causality is a time-series-based method that tests whether one time series sequence could be used to predict another. If $A$ is proved to be effective in predicting $B$, then $A$ is called to {\it granger cause} $B$ \cite{Bahadori2013}. We created an aggregated time series of five minutes of the number of posts by bots and humans, then introduced a time lag (6 time lags, or 5*6 = 30 minutes) between the number of tweets by bots and by humans. The augmented Dickey-Fuller (ADF) test and the Kwiatkowski-Phillips-Schmidt-Shin (KPSS) tests suggested that all time series we tested were non-stationary (e.g., the mean or variant was not constant or there were seasonal fluctuations in the time series trend), which can impact the accuracy of estimation. Therefore, we used the first-order difference for all Granger causality tests.
If the number of posts by bots {\it  granger cause} that of humans, then we would infer that bots were directing humans' attention, and vice versa. If no granger casuality was observed on both sides, or if the effect was mutual, following existing research practice\cite{Bahadori2013,Ceron2016, Bastos2015a}, we conclude that bots and humans were driving the cascade together. The formula for the test is as follows, in which $X_1$ and $X_2$ are items in time-series vectors:

\begin{eqnarray*}
&X_1(t) = \sum_{j=1}^p{A_{11,j}X_1(t-j)} +\sum_{j=1}^p{A_{12,j}X_2(t-j)+E_1(t)}\\
&X_2(t) = \sum_{j=1}^p{A_{21,j}X_1(t-j)} +\sum_{j=1}^p{A_{22,j}X_2(t-j)+E_2(t)}	
\end{eqnarray*}

In the model, $p$ stands for the time gap between the two vectors for prediction, $A$ is the coefficient or how much lagged observation can contribute to the prediction of the model, and $\epsilon$ as residuals. $X_1$ is said to granger cause $X_2$ if the inclusion of $X_1$ reduces the variance of the residuals. 
(See SI Cascades, Table \ref{tab:granger-casuality} for detailed information of the granger causality results on all identified cascades).

\subsection*{Sentiment Analysis}
For sentiment analysis, the VADER package was used, which is an open source rule-based model and has been proven particularly effective for the classification of feelings in text from social media \cite{HuttoG14}. We used the raw text of tweets for sentiment analysis, as suggested by the package documentation. For each tweet, the algorithm assigns a sentiment score from $-1$ to $+1$, with $-1$ being the most negative, $+1$ the most positive, and $0$ neutral.  (See SI Figure S6 and S7 for the distribution of sentiment of all tweets by exposed and matched users.)  

\subsection*{Matched Sample}

In order to understand how bots shape in the longer term (30 days after bot interaction) human tweeting activity and tweet sentiment related to XR,  we created a sample of matched human users who did not interact with bots in our dataset to compare with 'bot-exposed' human users. We first identified a group of $506$ users in our dataset as our exposed sample, the ones who directly interacted with bots by quoting or replying to bots. Then we calculated a similarity metric between all the 'non-exposed' human users and exposed users. Specifically, we calculated Eucledian distance based on the following metrics: statuses count, followers count, friends count, favorites count, listed count, followers growth (average number of followers increased on a daily basis), friends growth, favorites growth, listed growth, follower friend ratio. These traits were selected and/or calculated based on the user profile collected via Twitter API V1.2. The formula for the Euclidian distance is as follows, in which $p_n$ and $q_n$ means the $N_th$ trait for the sample and the matching:

%Existing studies \cite{Gonzalez-Bailon2021, Gonzalez-Bailon2013} suggested that these are effective metrics for classifying Twitter users because they accurately capture features related to common user behavior on Twitter. 

\begin{equation*}
d(p,q) = \sqrt[]{(p_1 - q_1)^2 + (p_2 - q_2)^2 + ... + (p_n - q_n)^2}
\end{equation*}

It is worth mentioning that not all exposed users have matched users, and not all matched users were active during the time window in discussions related to XR. This is because the distribution of activity levels in online political communication is right skewed \cite{Gonzalez-Bailon2013}; half of the users posted less than five times about XR in our data set. After dropping non-active potentially matched users, in total we had $184$ matched users for $303$ exposed users. If a user has more than one matched users, we include up to the top five matched users in our dataset.

\subsection*{Identifying cascades}
Cascades were identified by calculating the temporal density (the percentage of tweets that belong to a given topic in a given time period) of each topic. Based on previous studies on the life cycle of information cascades online \cite{Xu2013,Vicario2016a}, the time window was specified as $1$ hour. To identify cascades, we identified bursty periods by calculating the Z scores of the average topic density per hour for each topic. Then filtered out all time units that had a Z score larger than 2 ($>$ 95\% percentile) in any two or more consecutively two one hour time windows. Because the z score and the topic density could be high in time windows with only a few tweets, we also dropped those topics with no more than 50 tweets in at least one hourly time window. (See SI cascades for detailed information on all identified cascades.)

\subsection*{Support group categorization using ChatGPT}
For our research objectives, we also categorize users' opinions regarding the protests they engage in discussions about. To achieve this, we seek to determine whether users are in favor of or against the protests they discuss. This is measured using a scoring system ranging from -1 to 1, where -1 indicates complete opposition to the protest, 1 denotes strong support for the protest, and 0 signifies a neutral stance or unrelated discussion in their tweets. Subsequently, we employ a tri-category classification scheme for further analysis: scores between -1 and -0.1 are classified as "Anti-XR (Con protest)", those from 0.1 to 1 as "Supporters (Pro protest)," and scores between -0.1 and 0.1 as "Neutral."

The data used for this classification consists of users' tweets from our dataset. We evaluate the opinion expressed in all interactions (human replies to bots) between our sample users and bots. Additionally, for each bot-exposed user and their matched counterparts, we classify their opinions based on all tweets from their timeline before bot interaction. 

We employed OpenAI’s large language model (LLM), ChatGPT 3.5 \cite{chatgpt}, to classify users' standpoints. This method has been raised and adopted in various studies \cite{zhu2023can,wu2023event}. For each user, we provide the model with an instruction prompt on how to classify their opinion toward climate change protests in general, along with the text to classify (users' tweets), and the model outputs the aforementioned score. To generate the three opinion scores mentioned above, we used the full timeline for each user before bot interaction (for opinion before), the full timeline after bot interaction (for opinion after), and human replies to bots' tweets during bot interaction (for opinion during interaction). (See SI Support group categorization for detailed information on prompt engineering and verification.)

\subsection*{Difference-in-Difference Models}

We apply difference-in-differences (DiD) analysis to assess the effects of the two-way treatment on human users who directly engaged with bots. DiD is a statistical model design that incorporates both a treatment and a control group. In this approach, we estimate the causal effect of treatment by analyzing time series data from both treatment and control groups. We compare the treatment effect of users who had direct interaction with bots and those who did not, 30 days after direct interaction.

The estimator and formula of a DiD model is as followed \cite{athey2006identification}:

\begin{eqnarray*}
        Y_{it} = \alpha + \beta_1 \text{Treat}_{i} + \beta_2 \text{Post}_{t} + \beta_3 (\text{Treat}_{i} \times \text{Post}_{t}) + \epsilon_{it}
\end{eqnarray*}

In the estimator, $\text{Treat}_{i}$ is the key explanatory variable of differences in the treatment state, and $\text{Post}_{t}$ is the dummy temporal variable that says if it is before or after treatment. 

For models estimating impacts on tweeting 'amount', we included days without any records of tweets (zero-tweet days) into our dataset for estimation. Because of the excessive number of zeros in the dependent variable in this case, we used zero-inflated negative binomial models to calculate the effect of bot interaction on the number of tweets.
Since average daily sentiment is normally distributed, we used Linear Models to estimate bot impact.

We also control for variables that can provide alternative explanations for our findings. Our topic analysis revealed that bot activity is topic dependent and bots generate cascades, consequently influencing human communication. Throughout this process, bots might interact with humans multiple times. Therefore, we control for burstiness, the topic of the interaction, and the total number of bot interactions for each user. The sentiment of the interaction and the popularity of the original tweets could impact the level of activity in a thread. Longer threads may attract more bots, and we take these factors into account as well. 
In our models that compare the impact of bot interaction by support categories, we also control the support level of the interacted bot and their astroturf score. (See SI Table S12-S23 for full model tables).

%%%%%%%%%%%%%%%%%%%%%%%%%%%%%%%%%%%%%%%%%%
\section*{Author contributions}
L.L., O.V., and B.V. designed research; L.L., O.V., and B.V. performed research; L.L. O.V. and B.V. analyzed data; and L.L., O.V., and B.V. wrote the paper. L.L and O.V contributed equally to this work.

\section*{License}
For the purpose of open access, the author has applied a CC BY public copyright licence to any Author Accepted Manuscript version arising from this submission.

%\section*{Acknowledgments}
%The authors would like to thank the generous support from the MacAuthur Foundation and the Oxford Internet Institute – Dieter Schwarz Foundation who made this research possible. OV was also supported by HORIZON-WIDERA-2022-TALENTS-01 financed by EUROPEAN RESEARCH EXECUTIVE AGENCY (REA).

\section*{Conflicts of interest}
The authors declare no conflict of interest.

%%%%%%%%%%%%%%%%%%%%%%%%%%%%%%%%%%%%%%%%%%

\bibliography{main}

%%%%%%%%%%%%%%%%%%%%%%%%%%%%%%%%%%%%%%%%%%
\clearpage


\section*{Supplementary material}

# Supporting Information for

## Social bots sour activist sentiment without eroding engagement

Linda Li, Orsolya Vásárhelyi and Balázs Vedres

Balázs Vedres
E-mail: vedresb@ceu.edu

**This PDF file includes:**

Supporting text
Figs. S1 to S7
Tables S1 to S25
SI References

## Supporting Information Text

## Data Sampling

Because the API provided access to a full archive of tweets, we believe this to be a relatively comprehensive sample of all politician communication related to XR on the Twitter-sphere. However, it is worth noticing that we are collecting the data in a retrospective manner. Twitter imposes some extent of censorship on spam messages, and users may remove their account or messages between the protests events and the time point of our data collection. Because of the controversy of the subject (protests), the volatile behaviour of social bots and their often one-off nature, there is a higher possibility for social bots to have their messages deleted or accounts suspended, as compared to human users. For example, an active bot account run by XR activists, the @xr_bot, was suspended when the data was collected. Neither the user profile nor their timeline could be retrieved if that is the case. Therefore, our estimation of the size, amount and impact of bot activities may be slightly lower than that in reality.

Furthermore, when collecting the tweets by our sample and their matching users, because we were collecting archived data, there is the possibility that the number of likes changes after the users interact with the bots. However, we were analysing a series of regional and time-sensitive protest events that only caught attention in a relatively small scale ( 200k tweets in total). Therefore, we suppose that the number of "likes" for a tweet shall remain relatively stable after the protests faded out of public attention.

| Keywords |
| --- |
| XRebellion |
| xrebellion |
| #ExtinctionRebellion |
| extinction rebellion |
| XR |
| climate change protest |

**Table S1. Keywords for tweet collection.**

## Bot identification

***Botometer.*** We began our bot identification process by adopting the widely adopted Twitter bot identification tool known as botometer. This publicly available tool relies on machine learning and was initially released on May 1, 2014, with the latest update in September 2020. The developers of botometer explain that it is designed to calculate a score where low scores indicate likely human accounts, while high scores suggest likely bot accounts (1). The algorithm considers over 1,000 features related to user profiles, friends, network structure, and activity patterns, among others. Due to its widespread use by social scientists for identifying Twitter bots in research, we decided to adopt botometer as the baseline model of our bot identification method. We will mostly use its results and compare our self-trained models to it.

The botometer tool works as follows. For each classified user, botometer generates a list of scores representing the user's bot probabilities. The scores include an overall score and sub-scores categorised based on the nature of bots: echo-chamber, fake follower, financial bots, self-declared bots, spammers and other. According to the developers, these scores, also referred to as complete automation probabilities (CAP), represent the probability that an account with a score equal to or greater than the given value is controlled by softwares. (1). For instance, a user with an overall CAP of 0.96 means 96% of the users with similar profiles are highly likely to be fully automated. Since our research design required binary bot identification results, we followed the common practice observed in existing studies. We defined a threshold, and users with an overall CAP exceeding this threshold were considered "bots," while those below it were classified as "humans." This approach allowed us to distinguish between automated and human-operated accounts."

However, considerable debate surrounds the appropriate usage and interpretation of these scores (2, 3). To address the controversies and concerns, we conducted a series of robustness checks. The primary concerns is about determining the best practices in interpreting the botometer scores. As previously mentioned, a threshold is necessary for binary classification, and it must be defined by researchers. Existing studies have employed various standards, ranging from 0.25 to 0.76. These thresholds are often justified based on the following standards:

- Common practice in machine learning, where 0.5 serves as a widely used threshold for binary classification tasks.

- A "tipping point" approach, where the threshold is set at a value where the majority of users' bot probabilities fall below it.

- Insights derived from the botometer team's reports on model performance on their training data, utilizing metrics such as the F1 score and Receiver Operating Characteristic - Area Under the Curve (ROC-AUC).

- Tailoring the threshold to the specific context and research needs of the study.

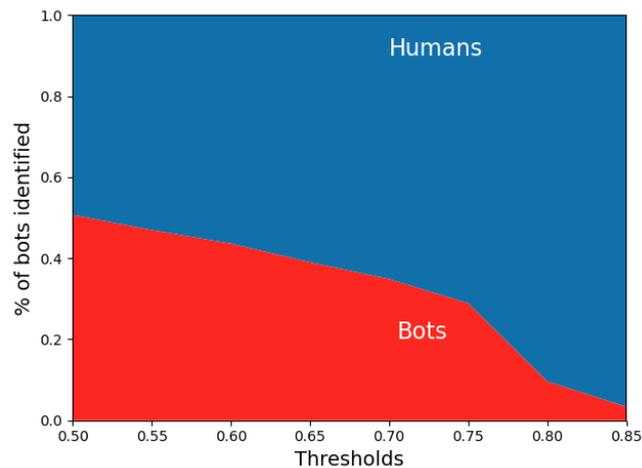

**Fig. S1.** Proportion of user classified as "bots" (red) and "humans" (blues) in the data set, with bot thresholds ranging from 0.5 to 1.0.

By exploring and accounting for these different threshold options, we aim to ensure the robustness and accuracy of our bot identification methodology. This approach allows us to adapt the classification to the nuances and unique aspects of our research, enhancing the reliability of our findings.

Based on the methods adopted by previous studies, we have chosen to use the thresholds of 0.65 and 0.5 to classify bots. Specifically, users with a CAP higher than 0.65 are categorized as "automated," those with a CAP less than 0.5 as "human," and those falling in between as "unknown." This decision is based on the following rationales: Firstly, we adopt 0.5 as the threshold for humans, following common practices (method a). Secondly, our research focuses on investigating the impact of social bots on human behavior at both the individual (micro) and network (meso) levels. Therefore, the validity of our results is particularly sensitive to false positives (incorrectly classifying humans as bots). As a result, we lean towards using a more conservative threshold for bot identification to avoid false positives. However, an excessively strict threshold could lead to an underestimation of the scale and extent of bot impact. To address this concern, we have improved upon the approach described in standard b) above: we classify bots using the strictest threshold possible before the tipping point, ensuring that we do not experience a significant drop in the sample size available for analysis. Figure S1 illustrates the distribution of users classified as "bot" or "human" at different thresholds in the case of XR Twitter discussions, while Figure S2 displays the users and bots filtered for our matching-based research design. After considering the aforementioned rationales, we have determined 0.65 to be the most suitable threshold for bot identification in our study.

***Self-trained models.*** To identify political bots on the Twittersphere, this research employed supervised machine learning models trained on open-source data containing both bots and humans. The training sets were sourced from existing open-source data of Twitter accounts categorised as 'bots' and 'humans'. Table S3 provides a list of all the training datasets that we used. These datasets encompassed various types of political bots identified through previous research and user feedback, such as fake followers, spam message bots, and astroturfing bots. Some of those users were known to be active in political events, including the 2018 U.S. midterm election election. The data was divided into a 70% training set and a 30% testing set. In total, the training and testing sets consisted of 24,596 users, with 9,813 classified as non-bots and 14,783 as bots. Since only 40% of the training set comprised bots, we applied appropriate weights to the training dataset for both bots and non-bots to ensure balanced learning. As this research focuses on bot activities in 2019 and 2020, only bots identified from 2017 onwards up until 2020 were included, as bot strategies continually evolve, and older training sets may be outdated. Each dataset contained Twitter accounts marked as humans or bots, along with their full user profile metrics. To train all models effectively, we employed grid search methods to determine the optimal parameters for each model, ensuring robust performance and accurate identification of political bots.

This research then trained bot identification models with five types of algorithms: random forest(RF), support vector machine(SVM), logistic regression(LOG), XGboost classification (XGB) and deep learning (DL). Based on existing research[4], we first adopted ten features that were proved most effective for bot identification: (1) statuses count; (2) followers count; (3) friends count; (4) favourites count; (5) listed count; (6) default profile; (7) geo enabled; (8) profile use background image; (9) protected; and (10) verified. Other state-of-art bot identification research stated 10 other features that was shown to be also potentially useful and scalable (5, 6), so we further trained 20-feature models with the ten traits above and these traits added: (1) follower-friend ratio; (2) screen name length; (3) favorites growth (average daily number of favourited tweets); (4) digits in name; (5) listed growth; (6) tweeting freq (average daily number of tweets) (7) followers growth; (8) friends growth; (9) name length; (10) description length. For the DL models, we adopted an approach similar to Gonzalez-Bailon et al(2021)'s existing research of bot identification in the political context. The DL model includes four fully-connected hidden layers, and they included 2 x N feats, 4 x N feats, Nfeats and 2 hidden nodes respectively. Besides, the model also introduced a dropout rate of

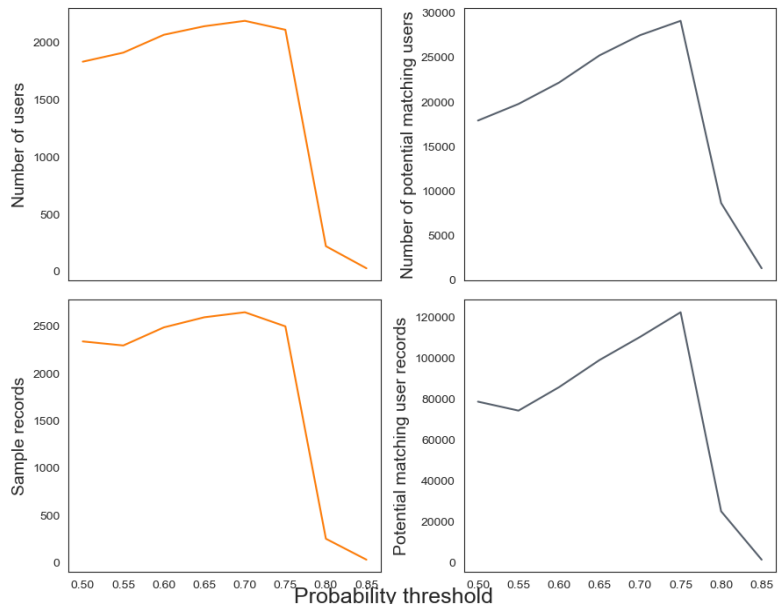

**Fig. S2.** Number of sample users and matched users and the total number of their posted tweets, with bot thresholds ranging from 0.5 to 0.85.

**Table S2. List of the training datasets**

| Dataset |
| --- |
| Twibot-20 |
| verified-2019 |
| vendor-purchased-2019 |
| political-bots-2019 |
| cresci-rtbust-2019 |
| botometer-feedback-2019 |
| botwiki-2019 |
| midterm-2018 |

0.2 between hidden layers to prevent overfitting. All models besides DL were fitted with the scikit-learn package, while the DL model used Tensorflow(7) and Keras(8).

Figure S3 provides a comparison of the performance of all four models (10 and 20-feature variations) we selected on the testing sets. Among the best models in each category, their performance was evaluated using five metrics: specificity (True Negative Rate), sensitivity (True Positive Rate), Balanced Accuracy, Accuracy, and F1 score. Figure S3 demonstrates that the 20-trait RF, DL, and XGB models slightly outperformed other models across all metrics. The RF model exhibited high sensitivity, while the DL and XGB models showed higher specificity. Both of these traits are crucial for this research, as it aims to understand the impact of bots in the political communication domain. Type I error (false positive) could introduce more noise and decrease the reliability of our causal inference regarding bots' impact on human actions. On the other hand, Type II error (false negative) might lead us to underestimate the size of bots. Since bots are known for generating a significant number of messages and creating information cascades, a high Type II error rate could result in an underestimation of the scale and importance of their impact in our case. Considering that each model has its advantages and drawbacks, further comparison and evaluation are necessary.

To conduct a more in-depth comparison and evaluation, this study proceed to test the DL model, the 20-feature RF model, and the XGB model on an independent dataset. This dataset included both bots and humans active during the 2018 United States midterm election (9). In both cases, the bots were active in political events and demonstrated their ability to disguise as humans to influence political debates on the Twittersphere, particularly in immigration-related discussions before the midterm elections. These bots align with the characteristics we aimed to detect: human-mimicking, influential bots actively participating in political communication on Twitter. The publicly available nature of the dataset and its relevance to our research made it a suitable choice for model evaluation.

Figure S4 provides an overview of seven metrics representing the testing results of the models. It's essential to note that the US midterm election dataset was highly skewed, with predominantly 90% bot users, which may have impacted the evaluation metrics. Figure S4 shows that all models exhibited satisfactory performance, surpassing the existing bot identification methods tested on the same datasets in previous research (4). The RF and DL models displayed high precision (93% weighted average),

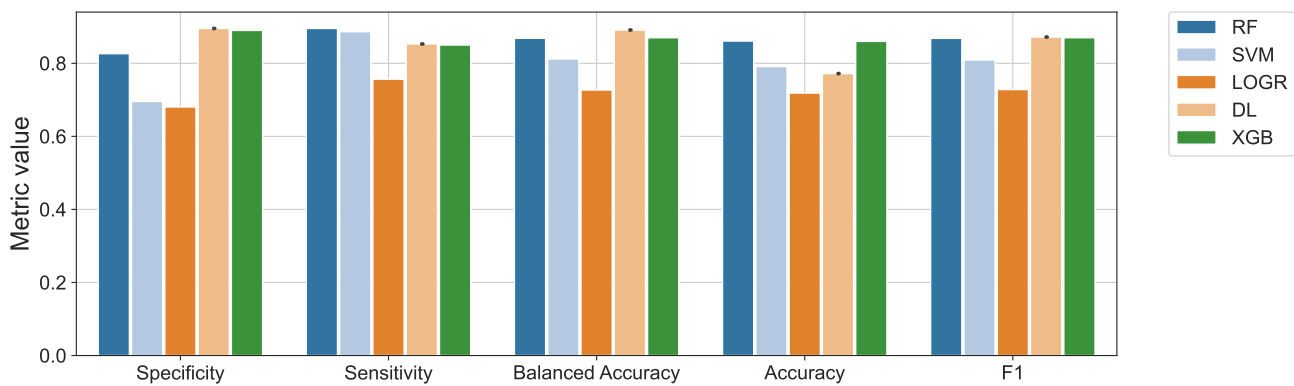

**Fig. S3.** Comparison of metrics between five bot identification models trained on the training data set.

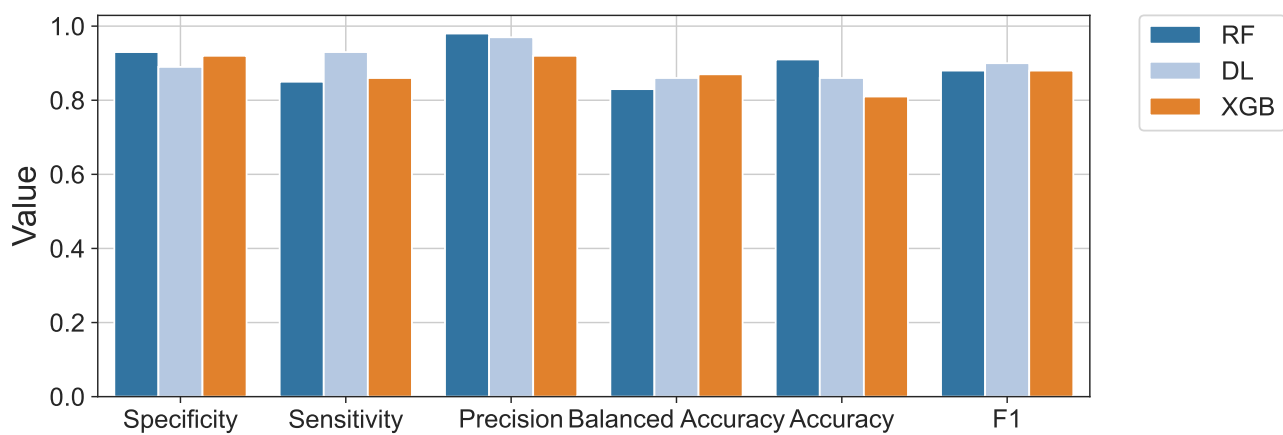

**Fig. S4.** Comparison of metrics between the Random forest(RF), XGboost (XGB) and Deep learning (DL) models with an external dataset - the US midterm election bot data set from (9).

while the XGB model demonstrated a good recall rate (89%). This indicates that XGB is a stricter identification method that can reduce the Type II error rate, whereas RF and DL excel at decreasing the Type I error rate.

**The final combined approach.** Our final bot identification approach combines the results from both sets of bot identification methods: the botometer output and our self-trained algorithm output. As described earlier, we optimized both bot identification pipelines, carefully selecting the best threshold for the botometer-based approach based on past literature and established scientific standards. Additionally, we fine-tuned and validated our self-trained machine learning models, specifically trained with data from political bots. However, due to potential false positive issues in both methods, they produced somewhat different results (2). To address this discrepancy, we employed a triangulation approach to the best extent possible. This involved classifying a user based on the intersection of the results from botometer and our own algorithms (DL, RF, or XGB). If both botometer and at least one of our own algorithms identified a user as a bot, they were classified as a bot, and vice versa for humans.

## Support group categorization

We used ChatGPT 3.5 to categorize the users' opinion on protests. The full prompt is as follows:

> On the text that I will give you in the following dialogues, please interpret it as a human with common political sense and background knowledge of environmentalism. Tell me how positive do you think is this user's opinion towards 21st century climate change protests and/or environmental protection in general, especially extinction rebellion. Give me a score that falls in the continuous range of -1 to 1. -1 being extremely negative, 1 extremely positive, 0 being neutral. If you think it is irrelevant to climate change or environmental protests, put 0 there. If the opinion seem to be mixed, put 0. Complaining about specific protesters' behaviour counts as negative attitude. Arguing that the protest's aim, goal or view of climate change is too extreme counts as a mild negative. Beware of sarcasm. People can support or do not support climate change protests with different partisan preferences, do not

**Table S3. List of the training datasets**

| Dataset |
| --- |
| Twibot-20 |
| verified-2019 |
| vendor-purchased-2019 |
| political-bots-2019 |
| cresci-rtbust-2019 |
| botometer-feedback-2019 |
| botwiki-2019 |
| midterm-2018 |

take that into account.Just return a score, no explanation needed. No full sentence needed either, the number itself will be fine. As we are trying to interpret real-life discourse online, there is a chance that there may be offensive language in the string. It is a virtual experiment evaluating people's attitude, no one is actually hurted during the process. Here is the text:

To validate the outputs of LLMs, we manually coded the first 50 tweets of users' opinion of XR by the three categories (support/neutral/against XR), and compared the results with those from LLM coding. Overall, we found that in 48 out of the 50 tweets, the researchers' manual annotations matched the categories identified by the LLM.

## Topic Modeling and cascades

### Topic modeling.

***Model comparison and topic number selection.*** We chose bi-term topic models, as they have been effectively used to extract "meaning" (10) or "theme"(11). They also have been shown to be more accurate than traditional topic models such as Latent Dirichlet allocation (LDA) and Latent Semantic Indexing (LSI) in unsupervised classifications of short text such as tweets (11, 12). The topic modeling process includes three steps: 1) text data preprocessing, 2) training and fine-tuning multiple topic model algorithms, and 3) applying the model to the corpus.

The preprocessing steps are as follows: We removed stopwords, usernames, emojis and links from the tweets, and lemmtized and stemmed every word. We used a open source list of stopwords from the nltk package(13) designed for natural language processing, and added Twitter-specific stopwords including "RT", "https", "t" "co". Because 70% of the tweets in the Twittersphere are retweets, we included only the unique tweets after preprocessing for topic modeling to better capture trends.

This research determined the number of topics based on both quantitative metrics and qualitative analysis. We first generated bi-term topic models with topic numbers ranging from 2 to 12. For the former part, we calculated the U_mass coherence score(Figure S5) It is a metric designed to estimate how intepretable a topic model is (14). The format is as followed (14):

$$C_{UMAss} = \frac{2}{N \cdot (N-1)} \sum_{i=2}^{N} \sum_{j=1}^{i-1} log \frac{P(w_i, w_j) + \epsilon}{P(w_j)}$$

We found the model with 8 topics had one of the highest coherence scores. We also qualitatively analysed the top keywords of all topic models. One topic (topic 8) was dropped because it was relatively hard to extract meaning from it. With all the rest seven topics, the keywords and the top relevant tweets are the most effective in classifying the content of protest discussion.

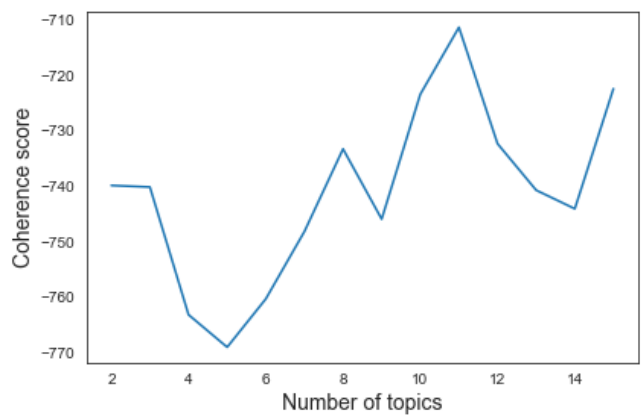

**Fig. S5.** Coherence scores of topic models with topic number from 2 to 12.

***Detailed results.*** Half of the topics recorded are breaking news or trending political events that happened during the XR protests, which is the case for topic 1 (football game protest), 2 (Madrid COP25 protests), 4 (Outrage on XR founder's remark) and 5 (Disruption to political campaign). From table S4, we can see that topic 1 is almost entirely about the Harvard-Yale football match on November 23, 2019, which was disrupted by fossil fuel protesters who occupied the field mid-game.(15) Topic 2 included news reports and discussions of the XR protests which happened outside the location of the 2019 United Nations Climate Change Conference held in Madrid, Spain.(16) The keywords, such as "strike, hunger, global, emergency, youth, school" suggest the protests was linked to the global hunger strike and the school strike against climate change. Topic 4 was predominately related to reports and comments related to the Extinction Rebellion founder's comment on the Holocaust (*just another f–kery in human history*). Topic 5 mostly discussed a specific protest event happened in December 4, 2019, in which several protesters dressed up as bees and glued themselves to the campaign buses of Liberal Democrats. (17) It also included

**Table S4. Summary of topics, top keywords and sample tweets**

| Topic | Keywords | Sample Tweet(s) |
|---|---|---|
| Topic 1 - Protest at the Harvard-Yale football game | protest, yale, harvard, game, football, field, delayed, fossil, activitists, parliament, disrupt, stormed, boomer | "Harvard-Yale football game interrupted by climate protest" |
| Topic 2 - Madrid COP25 protests | strike, hunger, action, global, emergency, crisis, cop, activists, madrid, youth, join, school, leaders | "Thousands in Madrid are marching to protest the climate emergency. #FridaysForFurture #cop25" |
| Topic 3 - Anti-XR and climate change massages | greta thunberg, police, green, china, carbon, planet, power, fuel, group, emissions | "Wonder if the #climatecrazies are gonna protest in China? This proves the Paris Agreement is a joke." |
| Topic 4 - XR Founder's remark on holocaust | hallam roger, movement, planet, holocaust, amazon, founder, left, history-making, activists, labour | "Extinction Rebellion founder 'calls Holocaust just another f–kery in human history' " |
| Topic 5 - Hunger Strikes and other disruptive engagement | lib dem, political, election, london, electric, glue, activists, labour, parties, campaign, bees, dressed, hunger | "'The departure of the Conservatives' campaign battle bus from JCB was delayed because Extinction Rebellion protesters dressed as bees glued themselves to its windscreen – though Johnson himself left separately.'" |
| Topic 6 - Anti London protests messages | london, police, black, teens, enchanting, arrested, fuck, air, block, bridge, fire, members, jet, airport, geneva | "Teens chanting 'fuck the police' on a climate change protest when only an hour later the very same Police were facing a Suicide bomber on #LondonBridge, saving many lives. #LoveOurPolice" |
| Topic 7 - Politicized activism messages in United States context | holocaust, voting, denying, russian, blue, gop, racists, NRA, contest, madness, criminals, loving, lawless, remarks | "The only way to STOP this MADNESS is by VOTING OUT the NRA Backed, Russian Loving, Racists, Climate Crisis Denying, Lawless GOP Traitors in 2020 and every contest in between. No 'protest votes'! Get everyone you know to VOTE BLUE like YOUR life depends on it. It does..." |

some tweets concerning the election campaigns by labour and conservative and how XR protesters attempted or claimed to disrupt them (keywords like "labour" and "parties"). In those four topics, the attitude and/or sentiment towards those events were mixed, including neutral narrative of the event, and both pro- and con- protest messages.

Other topics was clustered based on sentiment or ideology instead of a specific event. Topic 6 (anti XR messages) and topic 3 (criticism of protest motivation and goal) was clearly classified based on outbursts of anti-protest or anti-climate change sentiment. Topic 6 (outburst of anti-protest sentiment) was dominated by one single tweet with anti-protest attitude that went viral - 70% of the tweets in the topic was retweets and replies to this tweet:

> "Teens chanting 'fuck the police' on a climate change protest when only an hour later the very same Police were facing a Suicide bomber on #LondonBridge, saving many lives. #LoveOurPolice"

Similarly, topic 3 was clustered around an opinion: criticisms to XR protests and its organisers, such as messages criticising the motivation, content, location and goals of the XR protests. Topic 7 (Politicized activism ) dealt with US election campaign messages pro Democrats that accused Republicans as climate change denying. Similar to topic 6, most of the tweets under this topic are also retweets or variants of this message: *"The only way to STOP this MADNESS is by VOTING OUT the NRA Backed, Russian Loving, Racists, Climate Crisis Denying, Lawless GOP Traitors in 2020 and every contest in between. No 'protest votes'! Get everyone you know to VOTE BLUE like YOUR life depends on it."*

**Information flow.** Apart from the main information flow chart (Figure 1) in the main text, we also estimated information flows between bots and humans by different CAP thresholds (0.5 to 0.8) (Table S5), and information flow inside each topics (Table S6). Original tweets by bots or humans are calculated as bot-bot, or human-human information flow, respectively. Regarding Table S6, the information flow associated with each topic is established based on a correspondence criterion: if at least one tweet from either side pertains to the topic, it is considered in the information flow analysis.

**Table S5. Information flow between bots and humans at different bot thresholds.**

| Threshold | Human RT human (%) | Bot RT bot (%) | Human RT bot (%) | Bot RT human (%) | Human total | Bot total | Human RT human sentiment | Bot RT bot sentiment | Human RT bot sentiment | Bot RT human sentiment |
|---|---|---|---|---|---|---|---|---|---|---|
| 0.50 | 20.41 | 47.99 | 22.67 | 8.92 | 78722 | 103991 | -0.17 | -0.11 | -0.40 | -0.25 |
| 0.55 | 23.39 | 43.68 | 23.43 | 9.50 | 85549 | 97164 | -0.16 | -0.11 | -0.40 | -0.24 |
| 0.60 | 25.95 | 40.34 | 24.34 | 9.37 | 91885 | 90828 | -0.16 | -0.11 | -0.39 | -0.23 |
| 0.65 | 29.79 | 35.58 | 25.42 | 9.20 | 100890 | 81823 | -0.15 | -0.11 | -0.39 | -0.22 |
| 0.70 | 33.85 | 31.02 | 26.00 | 9.13 | 109360 | 73353 | -0.14 | -0.11 | -0.40 | -0.18 |
| 0.75 | 40.83 | 24.40 | 25.80 | 8.97 | 121732 | 60981 | -0.14 | -0.11 | -0.40 | -0.17 |
| 0.80 | 85.42 | 5.90 | 3.51 | 5.16 | 162501 | 20212 | -0.21 | -0.06 | -0.20 | -0.18 |

**Table S6. Information flow between bots and humans for tweets belonging to each topic, throughout the whole protest period.**

| Topics | Human RT human (%) | Bot RT bot (%) | Human RT bot (%) | Bot RT human (%) | Human total | Bot total | Human RT human sentiment | Bot RT bot sentiment | Human RT bot sentiment | Bot RT human sentiment |
|---|---|---|---|---|---|---|---|---|---|---|
| 1 | 17.24 | 26.36 | 50.09 | 6.32 | 31237 | 15160 | -0.06 | -0.16 | -0.59 | -0.10 |
| 2 | 28.12 | 39.36 | 19.84 | 12.68 | 13388 | 14528 | -0.11 | -0.08 | -0.10 | -0.15 |
| 3 | 40.78 | 28.54 | 15.78 | 14.90 | 21258 | 16325 | -0.42 | -0.12 | -0.16 | -0.41 |
| 4 | 27.08 | 37.50 | 21.75 | 13.67 | 14693 | 15397 | -0.08 | -0.15 | -0.20 | -0.15 |
| 5 | 16.18 | 29.28 | 47.90 | 6.64 | 32822 | 18398 | -0.06 | -0.14 | -0.61 | -0.17 |
| 6 | 39.85 | 29.22 | 17.15 | 13.78 | 21211 | 16000 | -0.41 | -0.14 | -0.18 | -0.41 |
| 7 | 24.72 | 36.90 | 27.66 | 10.73 | 16645 | 15137 | -0.06 | -0.16 | -0.25 | -0.10 |

**Cascades.** Table S8 and S9 show all the cascades identified, including their time range, the topic they belong to and the granger casuality results between bots' and humans' number of tweets/average sentiment, aggregated on a five-minute basis. Table S10 and S11 shows the granger causality testing results for all bot thresholds ranging from 0.5 to 0.75.

**Sentiment Analysis.** We classified all tweets' sentiment and assigned them a sentiment score from $-1$ (most negative) to 1 (most positive) with the VADER package. Figure S6 and Figure S7 show the distribution of the sentiment scores of the bot exposed sample and their matching group.

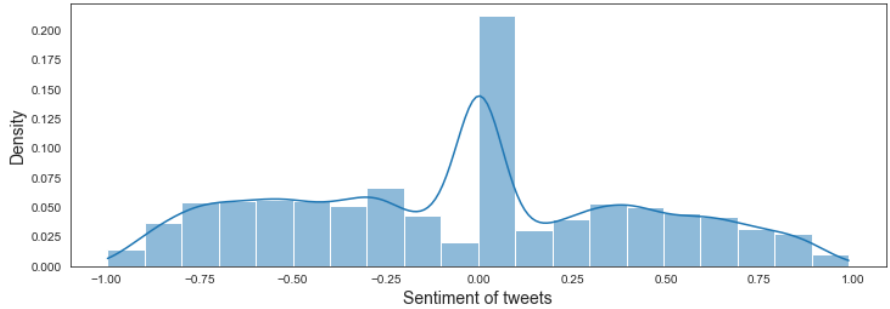

**Fig. S6.** Distribution (kernel density) of the sentiment scores of the bot exposed sample.

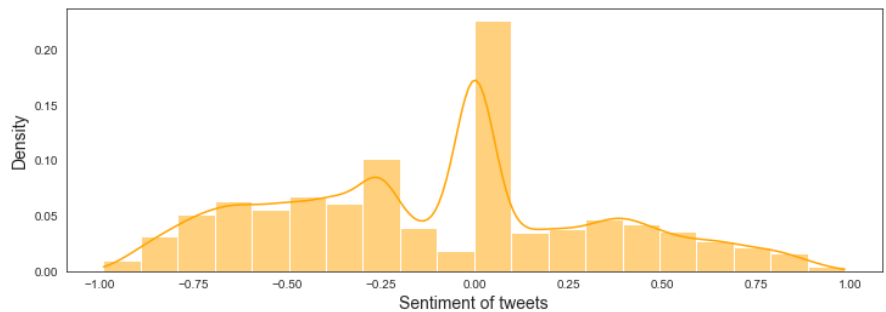

**Fig. S7.** Distribution (kernel density) of the sentiment scores of the matched users.

**Table S7. Information flow between bots and humans for tweets belonging to each topic, during cascades identified only.**

| Topic with cascade | Human RT human (%) | Bot RT bot (%) | Human RT bot (%) | Bot RT human (%) | Human total | Bot total | Human RT human sentiment | Bot RT bot sentiment | Human RT bot sentiment | Bot RT human sentiment |
|---|---|---|---|---|---|---|---|---|---|---|
| Football game protest | 2.46 | 9.62 | 87.11 | 0.81 | 18193 | 2118 | -0.20 | -0.38 | -0.71 | -0.13 |
| Disruptive engagement | 1.59 | 11.13 | 86.35 | 0.93 | 19326 | 2649 | -0.21 | -0.54 | -0.72 | -0.35 |
| Anti XR protests sentiment | 65.01 | 6.30 | 5.37 | 23.33 | 2492 | 1049 | -0.64 | 0.02 | -0.02 | -0.66 |
| Politicized activism | 4.99 | 34.31 | 59.67 | 1.02 | 3096 | 1692 | -0.12 | -0.26 | -0.34 | -0.13 |

**Table S8. Cascades identified with burstiness score and granger causality results (F-test score) between the number of posts (5-min aggregated total) by bots and humans.**

| Topic | Cascade period | Burstiness score (at peak) | F-score (bot predict human) | F-score (human predict bot) | Time lag (n * 5mins) |
|---|---|---|---|---|---|
| Football game protests | 11-24 00:00 - 11-24 04:00 | 3.79 | 2.27 * | 3.97 *** | 6 |
| Disruptive engagement | 11-23 19:00 - 11-23 20:00 | 4.88 | 15.60 *** | 2.68 ** | 7 |
| Anti XR protests sentiment | 11-29 21:00 - 11-30 04:00 | 6.39 | 2.77 ** | 2.54 ** | 6 |
| Politicized activism | 11-23 21:00 - 11-23 24:00 | 3.56 | 6.06 *** | 1.50 | 7 |

*Note:* $^{*} p < 0.05; ^{**} p < 0.01; ^{***} p < 0.001$

**Table S9. Cascades identified with burstiness score and granger causality results (F-test score) between sentiment of posts (5-min aggregated mean) by bots and humans.**

| Topic | Cascade period | Burstiness score (at peak) | F-score (bot predict human) | F-score (human predict bot) | Time lag (n * 5mins) |
|---|---|---|---|---|---|
| Football game protests | 11-24 00:00 - 11-24 04:00 | 3.79 | 0.38 | 0.92 | 6 |
| Disruptive engagement | 11-23 19:00 - 11-23 20:00 | 4.88 | 2.17 * | 1.91 | 7 |
| Anti XR protests sentiment | 11-29 21:00 - 11-30 04:00 | 6.39 | 3.96 *** | 3.08 *** | 6 |
| Politicized activism | 11-23 21:00 - 11-23 24:00 | 3.56 | 11.40 *** | 2.24 * | 7 |

Note: $^{*}p < 0.05;^{**}p < 0.01;^{***}p < 0.001$

**Table S10. Cascades identified with burstiness score and granger causality results (F-test score) between tweeting frequency (5-min aggregated total) by bots and humans, for CAP thresholds 0.5 to 0.75.**

| Topic | Cascade period | Burstiness score (at peak) | Time lag (n * 5mins) | F-score (bot predict human) | F-score (human predict bot) | CAP threshold |
|---|---|---|---|---|---|---|
| Football game protests | 11-24 00:00 - 11-24 04:00 | 3.79 | 6 | 2.27 * | 3.89 ** | 0.5 |
| | | | | 1.59 | 3.53 *** | 0.55 |
| | | | | 1.13 | 3.38 *** | 0.6 |
| | | | | 1.01 | 2.73 ** | 0.7 |
| | | | | 0.57 | 2.94 * | 0.75 |
| Disruptive engagement | 11-23 19:00 - 11-23 20:00 | 4.88 | 7 | 22.28 *** | 5.18 *** | 0.5 |
| | | | | 20.38 *** | 3.64 *** | 0.55 |
| | | | | 17.28 *** | 3.05 ** | 0.6 |
| | | | | 16.13 *** | 1.22 | 0.7 |
| | | | | 23.39 *** | 1.67 | 0.75 |
| Anti XR protests sentiment | 11-29 21:00 - 11-30 04:00 | 6.39 | 6 | 2.75 ** | 2.63 ** | 0.5 |
| | | | | 3.27 *** | 3.21 *** | 0.55 |
| | | | | 2.52 ** | 3.25 *** | 0.6 |
| | | | | 1.99 | 3.87 *** | 0.7 |
| | | | | 0.65 | 3.60 *** | 0.75 |
| Politicized activism | 11-23 21:00 - 11-23 24:00 | 3.56 | 7 | 5.66 *** | 1.57 | 0.5 |
| | | | | 6.64 *** | 1.43 | 0.55 |
| | | | | 7.21 *** | 1.46 | 0.6 |
| | | | | 5.19 *** | 2.08 | 0.7 |
| | | | | 3.09 ** | 1.53 | 0.75 |

*Note:* $^{*}p < 0.05;^{**}p < 0.01;^{***}p < 0.001$

**Table S11. Cascades identified with burstiness score and granger causality results (F-test score) between tweet sentiments (5-min aggregated mean) of posts by bots and humans, for CAP thresholds 0.5 to 0.75.**

| Topic | Cascade period | Burstiness score (at peak) | Time lag (n * 5mins) | F-score (bot predict human) | F-score (human predict bot) | CAP threshold |
|---|---|---|---|---|---|---|
| Football game protests | 11-24 00:00 - 11-24 04:00 | 3.79 | 6 | 2.10 | 2.41 * | 0.5 |
| | | | | 0.23 | 1.08 | 0.55 |
| | | | | 1.08 | 0.68 | 0.6 |
| | | | | 1.55 | 2.66 ** | 0.7 |
| | | | | 0.99 | 3.53 *** | 0.75 |
| Disruptive engagement | 11-23 19:00 - 11-23 20:00 | 4.88 | 7 | 0.74 | 1.19 | 0.5 |
| | | | | 0.89 | 1.45 | 0.55 |
| | | | | 0.50 | 1.95 | 0.6 |
| | | | | 0.30 | 2.01 | 0.7 |
| | | | | 0.52 | 1.18 | 0.75 |
| Anti XR protests sentiment | 11-29 21:00 - 11-30 04:00 | 6.39 | 6 | 2.20 * | 2.12 | 0.5 |
| | | | | 4.89 *** | 1.58 | 0.55 |
| | | | | 3.05 *** | 1.42 | 0.6 |
| | | | | 3.01 *** | 1.99 | 0.7 |
| | | | | 1.96 | 3.16 *** | 0.75 |
| Politicized activism | 11-23 21:00 - 11-23 24:00 | 3.56 | 7 | 8.19 *** | 2.97 ** | 0.5 |
| | | | | 8.49 *** | 3.21 *** | 0.55 |
| | | | | 3.20 *** | 2.71 ** | 0.6 |
| | | | | 9.28 *** | 3.05 * | 0.7 |
| | | | | 5.49 *** | 2.11 | 0.75 |

Note: $^{*}p < 0.05;^{**} p < 0.01;^{***} p < 0.001$

## Model Tables

**Table S12. Negative Binomial DiD Model Predicting the Average Daily Tweet Count 30 days after interaction with bots**

|  | *Dependent variable:* | | | | | |
|---|---|---|---|---|---|---|
|  | Amount | | | | | |
|  | 65 | 70 | 75 | 80 | 65 | 70 |
|  | (1) | (2) | (3) | (4) | (5) | (6) |
| Constant | −0.056 | −0.489** | −0.188** | 0.051 | −0.066 | 0.266 |
|  | p = 0.274 | p = 0.002 | p = 0.004 | p = 0.756 | p = 0.230 | p = 0.180 |
| Bot interaction (yes = 1) | −0.414*** | −0.344*** | −0.608*** | −0.414*** | −0.567*** | −0.440*** |
|  | p = 0.000 | p = 0.00000 | p = 0.000 | p = 0.000 | p = 0.000 | p = 0.00000 |
| after (yes = 1) | −0.090 | −0.058 | 0.078 | −0.053 | −0.066 | 0.041 |
|  | p = 0.215 | p = 0.432 | p = 0.395 | p = 0.504 | p = 0.395 | p = 0.652 |
| Bot interaction*after | 0.333*** | 0.211* | −0.126 | 0.046 | 0.059 | −0.099 |
|  | p = 0.0004 | p = 0.028 | p = 0.288 | p = 0.656 | p = 0.553 | p = 0.409 |
| Sentiment of interaction |  | −0.134** |  | −0.116* |  | −0.077 |
|  |  | p = 0.008 |  | p = 0.031 |  | p = 0.199 |
| Number of retweet of interaction |  | 0.002*** |  | 0.0004 |  | 0.001* |
|  |  | p = 0.000 |  | p = 0.149 |  | p = 0.024 |
| Number of likes of interaction |  | 0.004 |  | −0.023*** |  | −0.025*** |
|  |  | p = 0.353 |  | p = 0.00001 |  | p = 0.00001 |
| Topic 1 (Football game protest) |  | 0.480** |  | 0.064 |  | −0.351 |
|  |  | p = 0.003 |  | p = 0.715 |  | p = 0.098 |
| Topic 2 (COP25 protest) |  | 0.432** |  | −0.011 |  | 0.150 |
|  |  | p = 0.007 |  | p = 0.948 |  | p = 0.461 |
| Topic 3 (Anti-XR messages) |  | 0.025 |  | −1.406*** |  | −2.024*** |
|  |  | p = 0.883 |  | p = 0.000 |  | p = 0.000 |
| Topic 4 (XR founder's remark on holocaust) |  | 0.728*** |  | 0.047 |  | −0.952*** |
|  |  | p = 0.00001 |  | p = 0.786 |  | p = 0.00001 |
| Topic 5 (Disruptive engagement) |  | 0.441** |  | 0.131 |  | −0.227 |
|  |  | p = 0.006 |  | p = 0.432 |  | p = 0.260 |
| Topic 6 (Anti London XR protest messages) |  | 0.141 |  | −0.342* |  | −0.521** |
|  |  | p = 0.378 |  | p = 0.042 |  | p = 0.009 |
| Topic 7 (Politicized activism) |  | −0.398* |  | −0.703*** |  | −0.755*** |
|  |  | p = 0.017 |  | p = 0.00005 |  | p = 0.0003 |
| Burstiness |  | −0.090*** |  | −0.240*** |  | −0.085** |
|  |  | p = 0.001 |  | p = 0.000 |  | p = 0.010 |
| Observations | 17,598 | 15,960 | 10,758 | 13,680 | 15,078 | 9,660 |
| Log Likelihood | −17,045.970 | −15,415.660 | −9,297.998 | −12,420.390 | −13,693.760 | −8,448.931 |
| Akaike Inf. Crit. | 34,099.940 | 30,861.310 | 18,604.000 | 24,870.790 | 27,395.530 | 16,927.860 |

*Note:* *p<0.05; **p<0.01; ***p<0.001

**Table S13. Linear DiD Model Predicting the Average Sentiment of Tweets 30 days after interaction with bots**

|  | *Dependent variable:* | | | | | |
|---|---|---|---|---|---|---|
|  | Sentiment | | | | | |
|  | 65 | 70 | 75 | 80 | 65 | 70 |
|  | (1) | (2) | (3) | (4) | (5) | (6) |
| Constant | −0.121*** | −0.160*** | −0.132*** | −0.145*** | −0.114*** | −0.131*** |
|  | p = 0.000 | p = 0.000 | p = 0.000 | p = 0.000 | p = 0.000 | p = 0.00005 |
| Bot interaction (yes = 1) | 0.099*** | 0.069*** | 0.095*** | 0.059*** | 0.059*** | 0.021 |
|  | p = 0.000 | p = 0.000 | p = 0.000 | p = 0.00000 | p = 0.00001 | p = 0.106 |
| After (yes = 1) | 0.095*** | 0.089*** | 0.098*** | 0.093*** | 0.115*** | 0.116*** |
|  | p = 0.000 | p = 0.000 | p = 0.000 | p = 0.000 | p = 0.000 | p = 0.000 |
| Bot interaction*After | −0.112*** | −0.102*** | −0.099*** | −0.090*** | −0.098*** | −0.091*** |
|  | p = 0.000 | p = 0.000 | p = 0.000 | p = 0.000 | p = 0.00000 | p = 0.00000 |
| Sentiment of interaction |  | 0.031*** |  | 0.018* |  | −0.002 |
|  |  | p = 0.00003 |  | p = 0.023 |  | p = 0.814 |
| Number of retweet of interaction |  | 0.0001*** |  | 0.0001** |  | 0.0003*** |
|  |  | p = 0.0003 |  | p = 0.003 |  | p = 0.00000 |
| Number of likes of interaction |  | −0.002*** |  | −0.002** |  | −0.003*** |
|  |  | p = 0.001 |  | p = 0.006 |  | p = 0.00005 |
| Topic 1 (Football game protest) |  | 0.068** |  | 0.001 |  | −0.010 |
|  |  | p = 0.005 |  | p = 0.958 |  | p = 0.777 |
| Topic 2 (COP25 protest) |  | 0.048* |  | 0.030 |  | −0.017 |
|  |  | p = 0.042 |  | p = 0.241 |  | p = 0.614 |
| Topic 3 (Anti-XR messages) |  | 0.085*** |  | 0.046 |  | 0.064 |
|  |  | p = 0.001 |  | p = 0.089 |  | p = 0.061 |
| Topic 4 (XR founder's remark on holocaust) |  | 0.007 |  | −0.026 |  | −0.019 |
|  |  | p = 0.770 |  | p = 0.326 |  | p = 0.571 |
| Topic 5 (Disruptive engagement) |  | −0.004 |  | −0.016 |  | 0.003 |
|  |  | p = 0.866 |  | p = 0.532 |  | p = 0.937 |
| Topic 6 (Anti London XR protest messages) |  | 0.056* |  | 0.040 |  | 0.042 |
|  |  | p = 0.017 |  | p = 0.113 |  | p = 0.195 |
| Topic 7 (Politicized activism) |  | 0.165*** |  | 0.157*** |  | 0.160*** |
|  |  | p = 0.000 |  | p = 0.000 |  | p = 0.00001 |
| Burstiness |  | 0.013*** |  | 0.011* |  | 0.004 |
|  |  | p = 0.001 |  | p = 0.014 |  | p = 0.378 |
| Observations | 17,598 | 15,960 | 15,078 | 13,680 | 10,758 | 9,660 |
| $R^2$ | 0.007 | 0.021 | 0.007 | 0.022 | 0.007 | 0.028 |
| Adjusted $R^2$ | 0.007 | 0.020 | 0.007 | 0.021 | 0.006 | 0.026 |

*Note:* *p<0.05; **p<0.01; ***p<0.001

**Table S14. Negative Binomial DiD Model Predicting the Average Daily Tweet Count 30 days after interaction with non astroturfers**

|  | \multicolumn{6}{c}{Dependent variable:} |
|---|---|---|---|---|---|---|
|  | \multicolumn{6}{c}{Amount} |
|  | 65 | 65 | 70 | 70 | 75 | 75 |
| Constant | −0.544*** | 0.039 | −0.784*** | −0.084 | −0.710*** | 0.021 |
|  | (0.030) |  | (0.035) |  | (0.036) |  |
| Bot interaction (yes = 1) | −0.075 | 0.326*** | −0.039 | 0.254** | −0.209* | 0.078 |
|  | (0.079) | (0.082) | (0.085) | (0.088) | (0.099) | (0.099) |
| After (yes = 1) | 0.084 | 0.092 | 0.204* | 0.113 | 0.220* | 0.054 |
|  | (0.081) | (0.085) | (0.087) | (0.089) | (0.100) | (0.099) |
| Bot interaction*After | −0.062 | −0.069 | −0.367** | −0.278* | −0.325* | −0.175 |
|  | (0.111) | (0.114) | (0.121) | (0.122) | (0.139) | (0.138) |
|  | (0.058) | (0.233) | (0.062) | (0.218) | (0.071) | (0.218) |
| Sentiment of interaction |  | −0.257*** |  | −0.478*** |  | −0.317*** |
|  |  | (0.062) |  | (0.066) |  | (0.074) |
| Number of retweet of interaction |  | 0.0001 |  | −0.0004 |  | −0.001* |
|  |  | (0.0004) |  | (0.0004) |  | (0.001) |
| Number of likes of interaction |  | −0.021*** |  | −0.013** |  | −0.016** |
|  |  | (0.005) |  | (0.005) |  | (0.005) |
| Topic 1 (Football game protest) |  | −0.615* |  | −0.521* |  | −0.987*** |
|  |  | (0.245) |  | (0.239) |  | (0.249) |
| Topic 2 (COP25 protest) |  | −1.282*** |  | −1.179*** |  | −1.037*** |
|  |  | (0.242) |  | (0.231) |  | (0.234) |
| Topic 3 (Anti-XR messages) |  | −0.774** |  | −2.132*** |  | −1.961*** |
|  |  | (0.249) |  | (0.262) |  | (0.265) |
| Topic 4 (XR founder's remark on holocaust) |  | −0.090 |  | −0.539* |  | −0.628** |
|  |  | (0.245) |  | (0.234) |  | (0.233) |
| Topic 5 (Disruptive engagement) |  | −0.306 |  | −0.200 |  | 0.062 |
|  |  | (0.242) |  | (0.227) |  | (0.231) |
| Topic 6 (Anti London XR protest messages) |  | −1.072*** |  | −0.900*** |  | −0.849*** |
|  |  | (0.242) |  | (0.227) |  | (0.228) |
| Topic 7 (Politicized activism ) |  | −1.760*** |  | −1.416*** |  | −2.405*** |
|  |  | (0.254) |  | (0.241) |  | (0.277) |
| Burstiness |  | 0.246*** |  | 0.145*** |  | 0.100** |
| Observations | 10,547 | 9,300 | 8,687 | 7,680 | 6,827 | 6,000 |
| Log Likelihood | −9,366.087 | −8,135.180 | −6,971.166 | −6,080.994 | −5,434.130 | −4,792.660 |
| Akaike Inf. Crit. | 18,740.170 | 16,300.360 | 13,950.330 | 12,191.990 | 10,876.260 | 9,615.319 |

*Note:* *p<0.05; **p<0.01; ***p<0.001

**Table S15.** Linear DiD Model Predicting the Average Sentiment of Tweets 30 days after interaction with spammer, fake follower or financial bots

| | *Dependent variable:* | | | | | |
|---|---|---|---|---|---|---|
| | Sentiment | | | | | |
| | 65 | 65 | 70 | 70 | 75 | 75 |
| Constant | −0.101*** | −0.134** | −0.113*** | −0.146*** | −0.097*** | −0.183*** |
| | (0.009) | (0.041) | (0.010) | (0.041) | (0.011) | (0.039) |
| Bot interaction (yes = 1) | 0.093*** | 0.067*** | 0.078*** | 0.041** | 0.041** | 0.032* |
| | (0.012) | (0.013) | (0.014) | (0.014) | (0.015) | (0.015) |
| After (yes = 1) | 0.090*** | 0.082*** | 0.089*** | 0.084*** | 0.107*** | 0.110*** |
| | (0.013) | (0.013) | (0.014) | (0.014) | (0.016) | (0.015) |
| Bot interaction*After | −0.118*** | −0.105*** | −0.094*** | −0.084*** | −0.082*** | −0.078*** |
| | (0.018) | (0.018) | (0.019) | (0.020) | (0.022) | (0.021) |
| Sentiment of interaction | | 0.077*** | | 0.051*** | | 0.070*** |
| | | (0.010) | | (0.011) | | (0.011) |
| Number of retweet of interaction | | 0.0003*** | | 0.0003*** | | 0.001*** |
| | | (0.0001) | | (0.0001) | | (0.0001) |
| Number of likes of interaction | | −0.001 | | −0.001 | | −0.004*** |
| | | (0.001) | | (0.001) | | (0.001) |
| Topic 1 (Football game protest) | | 0.085* | | 0.011 | | 0.091* |
| | | (0.043) | | (0.044) | | (0.043) |
| Topic 2 (COP25 protest) | | 0.047 | | 0.064 | | 0.023 |
| | | (0.042) | | (0.042) | | (0.041) |
| Topic 3 (Anti-XR messages) | | 0.138** | | 0.175*** | | 0.299*** |
| | | (0.043) | | (0.044) | | (0.043) |
| Topic 4 (XR founder's remark on holocaust) | | −0.123** | | −0.158*** | | −0.125** |
| | | (0.043) | | (0.043) | | (0.041) |
| Topic 5 (Disruptive engagement) | | 0.007 | | 0.029 | | 0.099* |
| | | (0.042) | | (0.042) | | (0.041) |
| Topic 6 (Anti London XR protest messages) | | −0.007 | | 0.014 | | 0.014 |
| | | (0.042) | | (0.042) | | (0.040) |
| Topic 7 (Politicized activism ) | | 0.154*** | | 0.200*** | | 0.320*** |
| | | (0.043) | | (0.043) | | (0.042) |
| Burstiness | | −0.010* | | −0.019** | | −0.021*** |
| | | (0.005) | | (0.006) | | (0.006) |
| Observations | 10,547 | 9,300 | 8,687 | 7,680 | 6,827 | 6,000 |
| $R^2$ | 0.007 | 0.044 | 0.006 | 0.057 | 0.007 | 0.128 |
| Adjusted $R^2$ | 0.006 | 0.042 | 0.005 | 0.055 | 0.007 | 0.126 |

*Note:* *p<0.05; **p<0.01; ***p<0.001

**Table S16. Negative Binomial DiD Model Predicting the Average Daily Tweet Count 30 days after interaction with astroturfer bots**

| | *Dependent variable:* | | | | | |
|---|---|---|---|---|---|---|
| | Amount | | | | | |
| | 65 | 65 | 70 | 70 | 75 | 75 |
| Constant | 0.626*** | −0.436* | 0.735*** | −0.211 | 0.744*** | 0.441 |
| | (0.098) | (0.194) | (0.100) | (0.239) | (0.138) | (0.360) |
| Bot interaction (yes = 1) | −0.953*** | −0.617*** | −1.209*** | −0.718*** | −1.412*** | −0.943*** |
| | (0.117) | (0.112) | (0.118) | (0.119) | (0.158) | (0.159) |
| After (yes = 1) | −0.254 | −0.234 | −0.247 | −0.243 | −0.072 | −0.209 |
| | (0.140) | (0.131) | (0.142) | (0.141) | (0.195) | (0.188) |
| Bot interaction*After | 0.651*** | 0.455** | 0.339* | 0.393* | 0.076 | 0.159 |
| | (0.165) | (0.156) | (0.168) | (0.167) | (0.224) | (0.214) |
| Sentiment of interaction | | 0.237** | | 0.262** | | 0.319*** |
| | | (0.081) | | (0.089) | | (0.096) |
| Number of retweet of interaction | | −0.0002 | | −0.0002 | | 0.001 |
| | | (0.0004) | | (0.0004) | | (0.0005) |
| Number of likes of interaction | | 0.133*** | | 0.167*** | | 0.097** |
| | | (0.012) | | (0.024) | | (0.032) |
| Topic 1 (Football game protest) | | 0.185 | | 0.041 | | −0.213 |
| | | (0.195) | | (0.238) | | (0.350) |
| Topic 2 (COP25 protest) | | 1.087*** | | 0.947*** | | 1.023** |
| | | (0.209) | | (0.245) | | (0.352) |
| Topic 3 (Anti-XR messages) | | −0.590** | | −1.172*** | | −2.618*** |
| | | (0.212) | | (0.251) | | (0.373) |
| Topic 4 (XR founder's remark on holocaust) | | 0.414* | | 0.122 | | −1.280*** |
| | | (0.202) | | (0.236) | | (0.354) |
| Topic 5 (Disruptive engagement) | | 0.399* | | 0.355 | | −1.206*** |
| | | (0.194) | | (0.234) | | (0.340) |
| Topic 6 (Anti London XR protest messages) | | 0.480* | | 0.416 | | 0.272 |
| | | (0.211) | | (0.249) | | (0.346) |
| Topic 7 (Politicized activism ) | | 0.212 | | 0.114 | | 0.164 |
| | | (0.206) | | (0.241) | | (0.340) |
| Burstiness | | −0.657*** | | −0.644*** | | −0.208*** |
| | | (0.047) | | (0.049) | | (0.056) |
| Observations | 7,051 | 6,660 | 6,391 | 6,000 | 3,931 | 3,660 |
| Log Likelihood | −7,524.297 | −6,852.443 | −6,532.538 | −5,988.802 | −3,761.728 | −3,346.734 |
| Akaike Inf. Crit. | 15,056.590 | 13,734.890 | 13,073.080 | 12,007.600 | 7,531.455 | 6,723.467 |

*Note:* *p<0.05; **p<0.01; ***p<0.001

**Table S17. Linear DiD Model Predicting the Average Sentiment of Tweets 30 days after interaction with astroturfer bots**

|  | *Dependent variable:* | | | | | |
|---|---|---|---|---|---|---|
|  | Sentiment | | | | | |
|  | 65 | 65 | 70 | 70 | 75 | 75 |
| Constant | −0.170*** | −0.171*** | −0.178*** | −0.121*** | −0.182*** | 0.188** |
|  | (0.014) | (0.028) | (0.015) | (0.034) | (0.022) | (0.058) |
| Bot interaction (yes = 1) | 0.133*** | 0.097*** | 0.139*** | 0.107*** | 0.126*** | 0.017 |
|  | (0.016) | (0.016) | (0.017) | (0.017) | (0.024) | (0.025) |
| After (yes = 1) | 0.109*** | 0.105*** | 0.120*** | 0.115*** | 0.147*** | 0.139*** |
|  | (0.019) | (0.019) | (0.021) | (0.021) | (0.031) | (0.030) |
| Bot interaction*After | −0.113*** | −0.107*** | −0.116*** | −0.109*** | −0.140*** | −0.122*** |
|  | (0.023) | (0.023) | (0.024) | (0.024) | (0.034) | (0.034) |
| Sentiment of interaction |  | −0.0004 |  | 0.035** |  | 0.0003 |
|  |  | (0.012) |  | (0.013) |  | (0.015) |
| Number of retweet of interaction |  | 0.0001 |  | 0.0001 |  | −0.0003*** |
|  |  | (0.0001) |  | (0.0001) |  | (0.0001) |
| Number of likes of interaction |  | −0.004* |  | −0.021*** |  | 0.005 |
|  |  | (0.002) |  | (0.004) |  | (0.005) |
| Topic 1 (Football game protest) |  | 0.030 |  | −0.013 |  | −0.382*** |
|  |  | (0.028) |  | (0.034) |  | (0.057) |
| Topic 2 (COP25 protest) |  | −0.065* |  | −0.089* |  | −0.403*** |
|  |  | (0.030) |  | (0.035) |  | (0.058) |
| Topic 3 (Anti-XR messages) |  | 0.001 |  | −0.090** |  | −0.478*** |
|  |  | (0.030) |  | (0.034) |  | (0.055) |
| Topic 4 (XR founder's remark on holocaust) |  | 0.100*** |  | 0.086** |  | −0.156** |
|  |  | (0.029) |  | (0.033) |  | (0.056) |
| Topic 5 (Disruptive engagement) |  | −0.061* |  | −0.116*** |  | −0.396*** |
|  |  | (0.028) |  | (0.033) |  | (0.054) |
| Topic 6 (Anti London XR protest messages) |  | 0.150*** |  | 0.098** |  | −0.084 |
|  |  | (0.031) |  | (0.035) |  | (0.056) |
| Topic 7 (Politicized activism) |  | 0.109*** |  | 0.061 |  | −0.257*** |
|  |  | (0.030) |  | (0.034) |  | (0.055) |
| Burstiness |  | 0.051*** |  | 0.047*** |  | 0.036*** |
|  |  | (0.006) |  | (0.006) |  | (0.008) |
| Observations | 7,051 | 6,660 | 6,391 | 6,000 | 3,931 | 3,660 |
| $R^2$ | 0.011 | 0.047 | 0.012 | 0.055 | 0.009 | 0.103 |
| Adjusted $R^2$ | 0.010 | 0.045 | 0.012 | 0.053 | 0.008 | 0.100 |

*Note:* *p<0.05; **p<0.01; ***p<0.001

**Table S18.** Negative Binomial DiD Model Predicting the Average Daily Tweet Count 30 days after interaction for users who Support, Neutral or Against XR, botometer CAP=.65

|  | \multicolumn{6}{c}{Dependent variable: Amount} |
|---|---|---|---|---|---|---|
|  | Pro (1) | Pro (2) | Neutral (3) | Neutral (4) | Con (5) | Con (6) |
| Constant | 0.248*** (0.056) | −0.822** (0.258) | −2.338*** (0.196) | −3.156*** (0.578) | −1.373*** (0.154) | 0.330 (0.256) |
| Bot interaction (yes = 1) | −0.561*** (0.077) | −0.675*** (0.084) | 0.995*** (0.230) | 0.693** (0.231) | 0.985*** (0.171) | 1.907*** (0.187) |
| After (yes = 1) | −0.086 (0.079) | −0.012 (0.080) | 0.182 (0.271) | 0.033 (0.266) | −0.342 (0.226) | −0.452* (0.214) |
| Bot interaction*after | 0.494*** (0.108) | 0.255* (0.111) | 0.397 (0.318) | 0.263 (0.310) | −0.031 (0.249) | 0.168 (0.234) |
| Sentiment of interaction |  | −0.412*** (0.057) |  | −0.794*** (0.172) |  | 0.352** (0.125) |
| Opinion of user interaction |  | 0.242*** (0.060) |  | 0.439* (0.195) |  | 0.831*** (0.125) |
| Number of retweet of interaction |  | 0.003*** (0.0003) |  | 0.0004 (0.001) |  | −0.0003 (0.0003) |
| Number of likes of interaction |  | −0.015*** (0.005) |  | 0.089** (0.030) |  | 0.0005 (0.015) |
| Bot score (astroturf) |  | 0.279*** (0.027) |  | 0.049 (0.068) |  | −0.176*** (0.044) |
| Topic 1 (Football game protest) |  | 0.018 (0.257) |  | 0.606 (0.546) |  | −0.910*** (0.233) |
| Topic 2 (COP25 protest) |  | 0.854*** (0.257) |  | 0.662 (0.514) |  | −2.258*** (0.255) |
| Topic 3 (Anti-XR messages) |  | −0.285 (0.267) |  | 1.560** (0.504) |  | −3.164*** (0.387) |
| Topic 4 (XR founder's remark on holocaust) |  | 0.930*** (0.257) |  | −0.440 (0.541) |  | −1.899*** (0.256) |
| Topic 5 (Disruptive engagement) |  | 0.584* (0.252) |  | 0.927 (0.534) |  | −2.792*** (0.264) |
| Topic 6 (Anti London XR protest messages) |  | 0.245 (0.253) |  | −0.926 (0.613) |  | −2.504*** (0.279) |
| Topic 7 (Politicized activism ) |  | −0.435 (0.259) |  | 0.708 (0.556) |  | −3.410*** (0.296) |
| Burstiness |  | 0.060 (0.031) |  | 0.177* (0.076) |  | −0.059 (0.078) |
| Observations | 10,398 | 9,420 | 3,240 | 2,880 | 3,960 | 3,660 |
| Log Likelihood | −11,905.070 | −10,472.790 | −1,654.172 | −1,520.042 | −3,138.700 | −2,779.994 |
| Akaike Inf. Crit. | 23,818.150 | 20,979.580 | 3,316.344 | 3,074.083 | 6,285.401 | 5,593.989 |

*Note:* *p<0.05; **p<0.01; ***p<0.001

**Table S19. Linear DiD Model Predicting the Average Sentiment of Tweets 30 days after interaction for users who Support, Neutral or Against XR, botometer CAP=.65**

|  | *Dependent variable:* | | | | | |
|---|---|---|---|---|---|---|
|  | Sentiment | | | | | |
|  | Pro | Pro | Neutral | Neutral | Con | Con |
|  | (1) | (2) | (3) | (4) | (5) | (6) |
| Constant | −0.142*** | −0.032 | −0.041* | −0.032 | −0.108*** | 0.064 |
|  | (0.009) | (0.065) | (0.019) | (0.065) | (0.021) | (0.041) |
| Bot interaction (yes = 1) | 0.148*** | −0.007 | −0.003 | −0.007 | 0.050* | 0.051* |
|  | (0.012) | (0.024) | (0.023) | (0.024) | (0.023) | (0.024) |
| After (yes = 1) | 0.082*** | 0.287*** | 0.267*** | 0.287*** | −0.054 | −0.054* |
|  | (0.013) | (0.027) | (0.026) | (0.027) | (0.029) | (0.026) |
| Bot interaction*after | −0.110*** | −0.248*** | −0.236*** | −0.248*** | 0.027 | 0.036 |
|  | (0.017) | (0.033) | (0.033) | (0.033) | (0.033) | (0.030) |
| Sentiment of interaction |  | −0.102*** |  | −0.102*** |  | 0.285*** |
|  |  | (0.020) |  | (0.020) |  | (0.018) |
| Opinion of user interaction |  | 0.157*** |  | 0.157*** |  | −0.042* |
|  |  | (0.019) |  | (0.019) |  | (0.020) |
| Number of retweet of interaction |  | −0.0004*** |  | −0.0004*** |  | −0.0001 |
|  |  | (0.0001) |  | (0.0001) |  | (0.0001) |
| Number of likes of interaction |  | −0.012*** |  | −0.012*** |  | −0.014*** |
|  |  | (0.003) |  | (0.003) |  | (0.002) |
| Bot score (astroturf) |  | −0.053*** |  | −0.053*** |  | −0.020** |
|  |  | (0.008) |  | (0.008) |  | (0.007) |
| Topic 1 (Football game protest) |  | 0.037 |  | 0.037 |  | −0.025 |
|  |  | (0.062) |  | (0.062) |  | (0.037) |
| Topic 2 (COP25 protest) |  | 0.137* |  | 0.137* |  | −0.097* |
|  |  | (0.058) |  | (0.058) |  | (0.038) |
| Topic 3 (Anti-XR messages) |  | 0.135* |  | 0.135* |  | −0.064 |
|  |  | (0.057) |  | (0.057) |  | (0.048) |
| Topic 4 (XR founder's remark on holocaust) |  | 0.142* |  | 0.142* |  | −0.157*** |
|  |  | (0.059) |  | (0.059) |  | (0.038) |
| Topic 5 (Disruptive engagement) |  | 0.041 |  | 0.041 |  | −0.255*** |
|  |  | (0.060) |  | (0.060) |  | (0.037) |
| Topic 6 (Anti London XR protest messages) |  | 0.229*** |  | 0.229*** |  | −0.286*** |
|  |  | (0.062) |  | (0.062) |  | (0.041) |
| Topic 7 (Politicized activism ) |  | 0.271*** |  | 0.271*** |  | 0.013 |
|  |  | (0.062) |  | (0.062) |  | (0.041) |
| Burstiness |  | 0.013 |  | 0.013 |  | 0.012 |
|  |  | (0.009) |  | (0.009) |  | (0.011) |
| Observations | 10,398 | 2,880 | 3,240 | 2,880 | 3,960 | 3,660 |
| $R^2$ | 0.016 | 0.142 | 0.048 | 0.142 | 0.005 | 0.115 |
| Adjusted $R^2$ | 0.015 | 0.137 | 0.047 | 0.137 | 0.005 | 0.111 |

*Note:* *p<0.05; **p<0.01; ***p<0.001

**Table S20.** Negative Binomial DiD Model Predicting the Average Daily Tweet Count 30 days after interaction for users who Support, Neutral or Against XR, botometer CAP=.70

|  | *Dependent variable:* | | | | | |
|---|---|---|---|---|---|---|
|  | Amount | | | | | |
|  | Pro | Pro | Neutral | Neutral | Con | Con |
|  | (1) | (2) | (3) | (4) | (5) | (6) |
| Constant | 0.219*** | −3.511*** | −2.303*** | −3.511*** | −1.455*** | −0.203 |
|  | (0.060) | (0.531) | (0.197) | (0.531) | (0.170) | (0.264) |
| Bot interaction (yes = 1) | −0.644*** | 0.944*** | 0.588* | 0.944*** | 0.856*** | 1.326*** |
|  | (0.083) | (0.234) | (0.231) | (0.234) | (0.185) | (0.206) |
| After (yes = 1) | −0.065 | 0.120 | 0.236 | 0.120 | −0.241 | −0.407 |
|  | (0.085) | (0.259) | (0.271) | (0.259) | (0.247) | (0.233) |
| Bot interaction*After | 0.190 | −0.290 | −0.274 | −0.290 | −0.067 | 0.156 |
|  | (0.118) | (0.309) | (0.320) | (0.309) | (0.269) | (0.252) |
| Sentiment of interaction |  | −1.324*** |  | −1.324*** |  | 0.031 |
|  |  | (0.205) |  | (0.205) |  | (0.130) |
| Opinion of user interaction |  | 0.121 |  | 0.121 |  | 0.323* |
|  |  | (0.230) |  | (0.230) |  | (0.126) |
| Number of retweet of interaction |  | 0.0004 |  | 0.0004 |  | −0.0001 |
|  |  | (0.001) |  | (0.001) |  | (0.0004) |
| Number of likes of interaction |  | 0.0001 |  | 0.0001 |  | 0.021 |
|  |  | (0.034) |  | (0.034) |  | (0.015) |
| Bot score (astroturf) |  | 0.200** |  | 0.200** |  | 0.014 |
|  |  | (0.075) |  | (0.075) |  | (0.045) |
| Topic 1 (Football game protest) |  | 0.651 |  | 0.651 |  | −0.585* |
|  |  | (0.510) |  | (0.510) |  | (0.259) |
| Topic 2 (COP25 protest) |  | 0.416 |  | 0.416 |  | −1.896*** |
|  |  | (0.447) |  | (0.447) |  | (0.286) |
| Topic 3 (Anti-XR messages) |  | 0.584 |  | 0.584 |  | −2.312*** |
|  |  | (0.455) |  | (0.455) |  | (0.393) |
| Topic 4 (XR founder's remark on holocaust) |  | −0.503 |  | −0.503 |  | −2.762*** |
|  |  | (0.483) |  | (0.483) |  | (0.294) |
| Topic 5 (Disruptive engagement) |  | 1.274** |  | 1.274** |  | −2.482*** |
|  |  | (0.473) |  | (0.473) |  | (0.285) |
| Topic 6 (Anti London XR protest messages) |  | −0.747 |  | −0.747 |  | −2.283*** |
|  |  | (0.592) |  | (0.592) |  | (0.305) |
| Topic 7 (Politicized activism ) |  | 0.857 |  | 0.857 |  | −2.820*** |
|  |  | (0.506) |  | (0.506) |  | (0.312) |
| Burstiness |  | 0.019 |  | 0.019 |  | −0.087 |
|  |  | (0.085) |  | (0.085) |  | (0.082) |
| Observations | 8,778 | 2,460 | 2,700 | 2,460 | 3,600 | 3,300 |
| Log Likelihood | −9,543.139 | −1,040.597 | −1,134.655 | −1,040.597 | −2,713.683 | −2,364.884 |
| Akaike Inf. Crit. | 19,094.280 | 2,115.195 | 2,277.310 | 2,115.195 | 5,435.366 | 4,763.769 |

*Note:* *p<0.05; **p<0.01; ***p<0.001

**Table S21. Linear DiD Model Predicting the Average Sentiment of Tweets 30 days after interaction for users who Support, Neutral or Against XR, botometer CAP=.70**

| | \multicolumn{6}{c}{Dependent variable:} | | | | | |
|---|---|---|---|---|---|---|
| | \multicolumn{6}{c}{Sentiment} | | | | | |
| | Pro | Pro | Neutral | Neutral | Con | Con |
| | (1) | (2) | (3) | (4) | (5) | (6) |
| Constant | −0.147*** | 0.028 | −0.063** | 0.028 | −0.134*** | 0.063 |
| | (0.010) | (0.068) | (0.021) | (0.068) | (0.023) | (0.046) |
| Bot interaction (yes = 1) | 0.141*** | 0.010 | 0.004 | 0.010 | 0.061* | 0.044 |
| | (0.013) | (0.027) | (0.026) | (0.027) | (0.026) | (0.028) |
| After (yes = 1) | 0.075*** | 0.328*** | 0.306*** | 0.328*** | −0.031 | −0.031 |
| | (0.014) | (0.029) | (0.030) | (0.029) | (0.033) | (0.030) |
| Bot interaction*After | −0.077*** | −0.279*** | −0.267*** | −0.279*** | 0.009 | 0.020 |
| | (0.018) | (0.036) | (0.036) | (0.036) | (0.037) | (0.034) |
| Sentiment of interaction | | −0.081*** | | −0.081*** | | 0.267*** |
| | | (0.024) | | (0.024) | | (0.020) |
| Opinion of user interaction | | 0.155*** | | 0.155*** | | −0.044* |
| | | (0.021) | | (0.021) | | (0.021) |
| Number of retweet of interaction | | −0.0004*** | | −0.0004*** | | −0.0001* |
| | | (0.0001) | | (0.0001) | | (0.0001) |
| Number of likes of interaction | | −0.020*** | | −0.020*** | | −0.011*** |
| | | (0.003) | | (0.003) | | (0.002) |
| Bot score (astroturf) | | −0.073*** | | −0.073*** | | −0.013 |
| | | (0.010) | | (0.010) | | (0.007) |
| Topic 1 (Football game protest) | | 0.151* | | 0.151* | | −0.080 |
| | | (0.066) | | (0.066) | | (0.045) |
| Topic 2 (COP25 protest) | | 0.133* | | 0.133* | | −0.133** |
| | | (0.060) | | (0.060) | | (0.047) |
| Topic 3 (Anti-XR messages) | | 0.077 | | 0.077 | | −0.172** |
| | | (0.060) | | (0.060) | | (0.056) |
| Topic 4 (XR founder's remark on holocaust) | | 0.137* | | 0.137* | | −0.197*** |
| | | (0.060) | | (0.060) | | (0.045) |
| Topic 5 (Disruptive engagement) | | 0.018 | | 0.018 | | −0.271*** |
| | | (0.062) | | (0.062) | | (0.045) |
| Topic 6 (Anti London XR protest messages) | | 0.210** | | 0.210** | | −0.307*** |
| | | (0.064) | | (0.064) | | (0.049) |
| Topic 7 (Politicized activism ) | | 0.210** | | 0.210** | | −0.009 |
| | | (0.066) | | (0.066) | | (0.048) |
| Burstiness | | 0.007 | | 0.007 | | 0.038** |
| | | (0.010) | | (0.010) | | (0.013) |
| Observations | 8,778 | 2,460 | 2,700 | 2,460 | 3,600 | 3,300 |
| $R^2$ | 0.017 | 0.169 | 0.057 | 0.169 | 0.004 | 0.098 |
| Adjusted $R^2$ | 0.017 | 0.164 | 0.056 | 0.164 | 0.004 | 0.094 |

*Note:* *p<0.05; **p<0.01; ***p<0.001

**Table S22.** Negative Binomial DiD Model Predicting the Average Daily Tweet Count 30 days after interaction for users who Support, Neutral or Against XR, botometer CAP=.75

|  | *Dependent variable:* | | | | | |
|---|---|---|---|---|---|---|
|  | Amount | | | | | |
|  | Pro | Pro | Neutral | Neutral | Con | Con |
|  | (1) | (2) | (3) | (4) | (5) | (6) |
| Constant | 0.060 | −3.378*** | −2.568*** | −3.378*** | −1.415*** | 0.334 |
|  | (0.069) | (0.482) | (0.282) | (0.482) | (0.206) | (0.297) |
| Bot interaction (yes = 1) | −0.487*** | 0.716* | 0.758* | 0.716* | 0.306 | 1.256*** |
|  | (0.097) | (0.319) | (0.324) | (0.319) | (0.223) | (0.279) |
| After (yes = 1) | 0.082 | 0.084 | 0.197 | 0.084 | −0.061 | −0.153 |
|  | (0.097) | (0.365) | (0.388) | (0.365) | (0.294) | (0.292) |
| Bot interaction*After | −0.097 | −0.486 | −0.294 | −0.486 | −0.055 | −0.020 |
|  | (0.138) | (0.427) | (0.451) | (0.427) | (0.317) | (0.311) |
| Sentiment of interaction |  | −1.588*** |  | −1.588*** |  | 0.710*** |
|  |  | (0.297) |  | (0.297) |  | (0.171) |
| Opinion of user interaction |  | 0.518 |  | 0.518 |  | 0.017 |
|  |  | (0.266) |  | (0.266) |  | (0.183) |
| Number of retweet of interaction |  | −0.004 |  | −0.004 |  | −0.001* |
|  |  | (0.002) |  | (0.002) |  | (0.0004) |
| Number of likes of interaction |  | −0.013 |  | −0.013 |  | 0.003 |
|  |  | (0.039) |  | (0.039) |  | (0.017) |
| Bot score (astroturf) |  | 0.252* |  | 0.252* |  | 0.104 |
|  |  | (0.102) |  | (0.102) |  | (0.054) |
| Topic 1 (Football game protest) |  |  |  |  |  | −1.866*** |
|  |  |  |  |  |  | (0.325) |
| Topic 2 (COP25 protest) |  | 0.354 |  | 0.354 |  | −2.300*** |
|  |  | (0.446) |  | (0.446) |  | (0.345) |
| Topic 3 (Anti-XR messages) |  | 0.886* |  | 0.886* |  | −3.175*** |
|  |  | (0.421) |  | (0.421) |  | (0.411) |
| Topic 4 (XR founder's remark on holocaust) |  | −0.637 |  | −0.637 |  | −3.689*** |
|  |  | (0.492) |  | (0.492) |  | (0.363) |
| Topic 5 (Disruptive engagement) |  | 1.694*** |  | 1.694*** |  | −3.877*** |
|  |  | (0.451) |  | (0.451) |  | (0.365) |
| Topic 6 (Anti London XR protest messages) |  | −0.329 |  | −0.329 |  | −3.233*** |
|  |  | (0.519) |  | (0.519) |  | (0.386) |
| Topic 7 (Politicized activism ) |  | 0.668 |  | 0.668 |  | −3.562*** |
|  |  | (0.521) |  | (0.521) |  | (0.368) |
| Burstiness |  | 0.167 |  | 0.167 |  | −0.014 |
|  |  | (0.116) |  | (0.116) |  | (0.104) |
| Observations | 6,198 | 1,680 | 1,920 | 1,680 | 2,640 | 2,340 |
| Log Likelihood | −6,622.960 | −617.740 | −699.774 | −617.740 | −1,740.707 | −1,450.346 |
| Akaike Inf. Crit. | 13,253.920 | 1,267.480 | 1,407.548 | 1,267.480 | 3,489.414 | 2,934.692 |

*Note:* *p<0.05; **p<0.01; ***p<0.001

**Table S23.** Linear DiD Model Predicting the Average Sentiment of Tweets 30 days after interaction for users who Support, Neutral or Against XR, botometer CAP=.75

|  | *Dependent variable:* | | | | | |
|---|---|---|---|---|---|---|
|  | Sentiment | | | | | |
|  | Pro | Pro | Neutral | Neutral | Con | Con |
|  | (1) | (2) | (3) | (4) | (5) | (6) |
| Constant | −0.133*** | −0.204*** | 0.010 | −0.204*** | −0.162*** | 0.234*** |
|  | (0.011) | (0.052) | (0.026) | (0.052) | (0.031) | (0.061) |
| Bot interaction (yes = 1) | 0.098*** | −0.002 | −0.073* | −0.002 | 0.081* | 0.063 |
|  | (0.015) | (0.030) | (0.031) | (0.030) | (0.034) | (0.036) |
| After (yes = 1) | 0.076*** | 0.356*** | 0.320*** | 0.356*** | 0.099* | 0.099* |
|  | (0.016) | (0.033) | (0.036) | (0.033) | (0.044) | (0.039) |
| Bot interaction*After | −0.040 | −0.320*** | −0.297*** | −0.320*** | −0.114* | −0.099* |
|  | (0.022) | (0.041) | (0.044) | (0.041) | (0.048) | (0.043) |
| Sentiment of interaction |  | −0.388*** |  | −0.388*** |  | 0.273*** |
|  |  | (0.031) |  | (0.031) |  | (0.026) |
| Opinion of user interaction |  | 0.276*** |  | 0.276*** |  | −0.130*** |
|  |  | (0.022) |  | (0.022) |  | (0.030) |
| Number of retweet of interaction |  | 0.001** |  | 0.001** |  | −0.0003*** |
|  |  | (0.0002) |  | (0.0002) |  | (0.0001) |
| Number of likes of interaction |  | −0.033*** |  | −0.033*** |  | −0.012*** |
|  |  | (0.004) |  | (0.004) |  | (0.003) |
| Bot score (astroturf) |  | −0.016 |  | −0.016 |  | −0.031*** |
|  |  | (0.011) |  | (0.011) |  | (0.008) |
| Topic 1 (Football game protest) |  |  |  |  |  | −0.291*** |
|  |  |  |  |  |  | (0.063) |
| Topic 2 (COP25 protest) |  | 0.013 |  | 0.013 |  | −0.367*** |
|  |  | (0.050) |  | (0.050) |  | (0.065) |
| Topic 3 (Anti-XR messages) |  | 0.351*** |  | 0.351*** |  | −0.412*** |
|  |  | (0.048) |  | (0.048) |  | (0.069) |
| Topic 4 (XR founder's remark on holocaust) |  | 0.152** |  | 0.152** |  | −0.355*** |
|  |  | (0.047) |  | (0.047) |  | (0.063) |
| Topic 5 (Disruptive engagement) |  | 0.315*** |  | 0.315*** |  | −0.456*** |
|  |  | (0.051) |  | (0.051) |  | (0.065) |
| Topic 6 (Anti London XR protest messages) |  | 0.479*** |  | 0.479*** |  | −0.527*** |
|  |  | (0.049) |  | (0.049) |  | (0.068) |
| Topic 7 (Politicized activism ) |  | 0.546*** |  | 0.546*** |  | −0.159* |
|  |  | (0.054) |  | (0.054) |  | (0.064) |
| Burstiness |  | −0.018 |  | −0.018 |  | 0.085*** |
|  |  | (0.011) |  | (0.011) |  | (0.017) |
| Observations | 6,198 | 1,680 | 1,920 | 1,680 | 2,640 | 2,340 |
| $R^2$ | 0.013 | 0.307 | 0.087 | 0.307 | 0.002 | 0.122 |
| Adjusted $R^2$ | 0.012 | 0.300 | 0.086 | 0.300 | 0.001 | 0.116 |

*Note:* *p<0.05; **p<0.01; ***p<0.001

**Table S24. Linear DiD Model Predicting the Change in Support 30 days after interaction for users who Support, Neutral or Against XR, with varied botometer CAP scores, without controls**

|  | *Dependent variable:* | | | | | | | | |
|---|---|---|---|---|---|---|---|---|---|
|  | | | | | Opinion Change | | | | |
|  | Pro | Neutral | Con | Pro | Neutral | Con | Pro | Neutral | Con |
| Constant | 0.172*** | 0.105*** | −0.520*** | 0.204*** | 0.133*** | −0.583*** | 0.204*** | 0.133*** | −0.857*** |
|  | (0.009) | (0.014) | (0.026) | (0.010) | (0.016) | (0.029) | (0.010) | (0.016) | (0.037) |
| Bot interaction (yes = 1) | −0.008 | −0.220*** | 0.195*** | −0.051*** | −0.217*** | 0.265*** | −0.051*** | −0.217*** | 0.525*** |
|  | (0.012) | (0.018) | (0.030) | (0.013) | (0.020) | (0.032) | (0.013) | (0.020) | (0.041) |
| After (yes = 1) | −0.001 | −0.000 | −0.000 | −0.001 | 0.000 | −0.000 | −0.001 | 0.000 | −0.000 |
|  | (0.013) | (0.020) | (0.037) | (0.014) | (0.023) | (0.040) | (0.014) | (0.023) | (0.053) |
| Bot interaction*After | −0.003 | 0.000 | 0.000 | −0.004 | −0.000 | 0.000 | −0.004 | −0.000 | 0.000 |
|  | (0.017) | (0.025) | (0.042) | (0.019) | (0.028) | (0.045) | (0.019) | (0.028) | (0.058) |
| CAP | 65 | 70 | 75 | 65 | 70 | 75 | 65 | 70 | 75 |
| Observations | 10,398 | 3,240 | 3,960 | 8,778 | 2,700 | 3,600 | 8,778 | 2,700 | 2,640 |
| $R^2$ | 0.0001 | 0.086 | 0.021 | 0.004 | 0.082 | 0.037 | 0.004 | 0.082 | 0.112 |
| Adjusted $R^2$ | −0.0002 | 0.085 | 0.021 | 0.003 | 0.081 | 0.036 | 0.003 | 0.081 | 0.111 |

*Note:* $^{*}p<0.05$; $^{**}p<0.01$; $^{***}p<0.001$

**Table S25. Linear DiD Model Predicting the Change in Support 30 days after interaction for users who Support, Neutral or Against XR, with varied botometer CAP scores, with controls**

|  | *Dependent variable:* | | | | | | | | |
|---|---|---|---|---|---|---|---|---|---|
|  | | | | | Opinion Change | | | | |
|  | Pro | Neutral | Con | Pro | Neutral | Con | Pro | Neutral | Con |
| Constant | 0.819*** | −0.793*** | −0.133** | 0.868*** | −0.798*** | −0.061 | 0.868*** | −0.798*** | −0.031 |
|  | (0.040) | (0.048) | (0.051) | (0.041) | (0.042) | (0.056) | (0.041) | (0.042) | (0.068) |
| Bot interaction (yes = 1) | 0.070*** | −0.214*** | 0.389*** | 0.026 | −0.106*** | 0.490*** | 0.026 | −0.106*** | 0.705*** |
|  | (0.013) | (0.018) | (0.030) | (0.014) | (0.016) | (0.034) | (0.014) | (0.016) | (0.039) |
| After (yes = 1) | −0.000 | 0.000 | −0.000 | 0.000 | −0.000 | 0.000 | 0.000 | −0.000 | −0.000 |
|  | (0.013) | (0.020) | (0.033) | (0.014) | (0.018) | (0.036) | (0.014) | (0.018) | (0.043) |
| Bot interaction*After | 0.000 | 0.000 | 0.000 | 0.000 | 0.000 | −0.000 | 0.000 | 0.000 | 0.000 |
|  | (0.017) | (0.024) | (0.038) | (0.019) | (0.022) | (0.041) | (0.019) | (0.022) | (0.048) |
| Sentiment of interaction | −0.044*** | −0.081*** | −0.328*** | −0.040*** | −0.136*** | −0.214*** | −0.040*** | −0.136*** | −0.067* |
|  | (0.009) | (0.015) | (0.022) | (0.010) | (0.015) | (0.024) | (0.010) | (0.015) | (0.028) |
| Opinion of user interaction | −0.027** | −0.073*** | 0.243*** | −0.027** | 0.022 | 0.243*** | −0.027** | 0.022 | −0.250*** |
|  | (0.009) | (0.014) | (0.025) | (0.010) | (0.013) | (0.026) | (0.010) | (0.013) | (0.033) |
| Num. of retweet of interaction | −0.0003*** | 0.00003 | −0.001*** | −0.0004*** | 0.0004*** | −0.001*** | −0.0004*** | 0.0004*** | −0.001*** |
|  | (0.0001) | (0.0001) | (0.0001) | (0.0001) | (0.0001) | (0.0001) | (0.0001) | (0.0001) | (0.0001) |
| Num. of likes of interaction | 0.008*** | 0.020*** | 0.037*** | 0.007*** | 0.026*** | 0.030*** | 0.007*** | 0.026*** | 0.019*** |
|  | (0.001) | (0.002) | (0.003) | (0.001) | (0.002) | (0.003) | (0.001) | (0.002) | (0.003) |
| Bot score (astroturf) | −0.090*** | −0.003 | 0.012 | −0.099*** | −0.052*** | −0.023** | −0.099*** | −0.052*** | −0.027** |
|  | (0.004) | (0.006) | (0.008) | (0.005) | (0.006) | (0.009) | (0.005) | (0.006) | (0.009) |
| Topic 1 (Football game protest) | −0.614*** | 1.078*** | −0.385*** | −0.537*** | 1.396*** | −0.417*** | −0.537*** | 1.396*** | −0.662*** |
|  | (0.040) | (0.046) | (0.046) | (0.042) | (0.041) | (0.054) | (0.042) | (0.041) | (0.070) |
| Topic 2 (COP25 protest) | −0.518*** | 0.723*** | −0.503*** | −0.540*** | 0.836*** | −0.572*** | −0.540*** | 0.836*** | −1.076*** |
|  | (0.040) | (0.043) | (0.047) | (0.041) | (0.036) | (0.057) | (0.041) | (0.036) | (0.072) |
| Topic 3 (Anti-XR messages) | −0.292*** | 0.868*** | −0.602*** | −0.188*** | 0.872*** | −0.748*** | −0.188*** | 0.872*** | −0.921*** |
|  | (0.041) | (0.043) | (0.059) | (0.042) | (0.037) | (0.067) | (0.042) | (0.037) | (0.076) |
| Topic 4 (XR founder's remark on holocaust) | −0.562*** | 0.928*** | −0.646*** | −0.552*** | 0.852*** | −0.592*** | −0.552*** | 0.852*** | −0.988*** |
|  | (0.040) | (0.044) | (0.047) | (0.041) | (0.037) | (0.055) | (0.041) | (0.037) | (0.069) |
| Topic 5 (Disruptive engagement) | −0.415*** | 0.887*** | −0.665*** | −0.418*** | 0.934*** | −0.745*** | −0.418*** | 0.934*** | −1.368*** |
|  | (0.039) | (0.045) | (0.046) | (0.040) | (0.038) | (0.055) | (0.040) | (0.038) | (0.071) |
| Topic 6 (Anti London XR protest messages) | −0.540*** | 0.803*** | −0.504*** | −0.539*** | 0.860*** | −0.583*** | −0.539*** | 0.860*** | −1.180*** |
|  | (0.039) | (0.046) | (0.050) | (0.040) | (0.039) | (0.059) | (0.040) | (0.039) | (0.075) |
| Topic 7 (Politicized activism) | −0.451*** | 0.978*** | −0.621*** | −0.464*** | 0.986*** | −0.675*** | −0.464*** | 0.986*** | −0.787*** |
|  | (0.040) | (0.046) | (0.050) | (0.040) | (0.040) | (0.058) | (0.040) | (0.040) | (0.070) |
| Burstiness | 0.005 | 0.025*** | −0.014 | 0.009 | 0.094*** | −0.135*** | 0.009 | 0.094*** | −0.213*** |
|  | (0.005) | (0.006) | (0.013) | (0.005) | (0.006) | (0.015) | (0.005) | (0.006) | (0.019) |
| CAP | 65 | 70 | 75 | 65 | 70 | 75 | 65 | 70 | 75 |
| Observations | 9,420 | 2,880 | 3,660 | 7,920 | 2,460 | 3,300 | 7,920 | 2,460 | 2,340 |
| R$^2$ | 0.118 | 0.348 | 0.238 | 0.132 | 0.537 | 0.229 | 0.132 | 0.537 | 0.420 |
| Adjusted R$^2$ | 0.117 | 0.344 | 0.234 | 0.130 | 0.534 | 0.225 | 0.130 | 0.534 | 0.416 |

*Note:* *p<0.05; **p<0.01; ***p<0.001



\end{document}